\documentclass{aa}  

\usepackage[varg]{txfonts}
\usepackage{graphicx}
\usepackage{natbib}
\bibpunct{(}{)}{;}{a}{}{,}

\begin{document}

   \title{HI shells in the Leiden/Argentina/Bonn HI survey}

   \author{S. Ehlerov\'a
          \and
          J. Palou\v{s}
          }

   \institute{Astronomical Institute, Academy of Sciences,
              Bo\v{c}n\'{\i} II 1401, Prague\\
              \email{ehlerova@ig.cas.cz}
             }

   \date{Received September 6, 2012; accepted November 9, 2012}

  \abstract
   {}
   {We analyse the all-sky Leiden/Argentina/Bonn HI survey, where we
    identify shells belonging to the Milky Way.}
   {We used an identification method based on the search of
    continuous regions of a low brightness temperature
    that are compatible with given properties of HI shells.}
   {We found 333 shells in the whole Galaxy. The size distribution
    of shells in the outer Galaxy
    is fitted by a power law with the coefficient of 2.6 corresponding to
    the index 1.8 in the distribution of energy sources. Their surface
    density decreases exponentially with a scale length
    of 2.8 $kpc$. The surface density of shells with radii $\geq 100\ pc$
    in the solar neighbourhood is $\sim 4\ kpc^{-2}$ and the 2D porosity is 
    $\sim 0.7$.}
   {}

   \keywords{ISM: bubbles -- Galaxy: structure -- radio lines: ISM}

   \maketitle

\section{Introduction}

The neutral interstellar medium (ISM) in galaxies is full of structures. 
Some of them are turbulent, wildly changing structures, others are more 
stable, while HI in particular shows very prominent structures. We see roughly 
elliptical objects with sizes from a few pc to more than 1 kpc, which may 
expand into the ambient medium. In the Milky Way they were identified for 
the first time by \citet{1979ApJ...229..533H} and since then they have been 
discovered in many external galaxies, both spiral and irregular, including the 
LMC and SMC 
\citep[for a more complete review see, e.g.,][]{2011AJ....141...23B}.

Shells may be the results of different processes in the ISM. It is a challenge 
to distinguish between shells created by the energy inserted by stellar 
winds, stellar radiation, and supernovae explosions; and structures created 
by other driving mechanisms, such as the infalling high-velocity clouds, 
gamma ray bursts, or the turbulence driven by galactic differential rotation. 
Walls of shells may develop Rayleigh-Taylor, Vishniac, Jeans, or other 
instabilities, thereby creating high-density and low-temperature regions, 
which become  sites of secondary triggered star formation. Dense 
condensations of cold gas along the edges of shells, secondary star formation, 
and compact HII regions of second-generation stars have been observed  by 
\citet{2006A&A...446..171Z}, \citet{2010A&A...523A...6D},
and others \citep[see also review by][]{2012IAUS..284..317E}.
However, many details on the stars vs. shells connection remain unknown, e.g., 
the number of shells per association, the energetics involved, and the 
fraction of shells with nonstellar origins.

Due to our position inside the Milky Way, observations of shells
in the Galaxy present special problems: unlike observations
of external galaxies, an observer has to process a substantial part, 
ideally all, of the sky to cover the whole Galaxy. The structures do not
lie at almost the same distance as in external galaxies, therefore results
are more dependent on the distance of objects from the Sun. 
Another difference is the shape of the shells perpendicular to the galactic 
symmetry plane. It is directly observable in the Milky Way, while not so 
clear in external galaxies, where we have to consider the orientation relative 
to the line of sight. In the Milky Way, there are many observations of 
individual shells and several papers dealing with the 
global picture \citep{1979ApJ...229..533H, 2002ApJ...578..176M, 
2007ApJ...661..285D, 2005A&A...437..101E}, but even these do not cover 
the whole Galaxy.

Since 2005 the all-sky Leiden/Argentina/Bonn HI survey has been available 
\citep[LAB;][]{2005A&A...440..775K}. 
Surveys of the sky available from the Parkes telescope 
\citep[GASS;][]{2010A&A...521A..17K} and the Effelsberg telescope 
\citep[EBHIS;][]{2011AN....332..637K}
exist or will exist in the near future and have a higher resolution than
LAB. There are other surveys with a higher resolution (The Southern Galactic
Plane Survey, the Canadian Galactic Plane Survey, the VLA Galactic Plane 
Survey), but they cover only the region along the Galactic equator. 

Identification of shells is usually done by eye. Human eyes 
are very good at connecting partially disconnected
parts of shells and using some software tools
\citep[e.g.some routines in KARMA][]{1997PASA...14..106G} may help. 
Still, an objective method is preferable. There are a
few attempts at an automatic identification process for shells: 
\citet{1998A&A...332..429T} compared a hydrodynamical model with observations; 
\citet{2007ApJ...661..285D} based their method on finding simple dynamical 
characteristics in HI data; and \citet{2005A&A...437..101E} (Paper 1) 
located continuous regions of low brightness temperature. The first attempt 
was not followed any further, probably because it is very difficult to 
construct a satisfactory and universal hydrodynamical model comparable 
to the real data. The second approach is aimed at small bubbles (typically 
$\sim 10\ \mathrm{pc}$) and was used for a portion of the Canadian Galactic 
Plane Survey data, which found $\sim 7000$ bubbles with radii 
$r_{\mathrm{sh}} < 40\ \mathrm{pc}$. The third approach is developed in 
this communication.

\section{Data and methods}

\subsection{Leiden-Argentine-Bonn survey}

The Leiden-Argentine-Bonn HI survey \citep[LAB,][]{2005A&A...440..775K} is an 
all-sky survey, a combination of the Leiden-Dwingeloo survey 
\citep{1997agnh.book.....H}
and the Instituto Argentino de Radioastronom\'{\i}a survey 
\citep{2005A&A...440..767B}, with angular resolution 36 arcmin 
(the pixel size 30 arcmin) and 
velocity resolution $1\ \mathrm{kms}^{-1}$ (the channel width is 
$1.0306\ \mathrm{kms}^{-1}$). We used the datacube limited in 
$v_{\mathrm{lsr}}$ to $(-250,+250)\ \mathrm{kms}^{-1}$.

\subsection{Identification process}

The identification process is divided into three steps: 
1) we trace {\it continuous regions around local minima} in 2D channel maps;
2) we construct {\it 3D structures} from 2D identifications from step 1
by connecting consecutive overlapping regions;
3) we eliminate structures that are not compatible with the given
   {\it properties of HI shells}.

\begin{figure*}
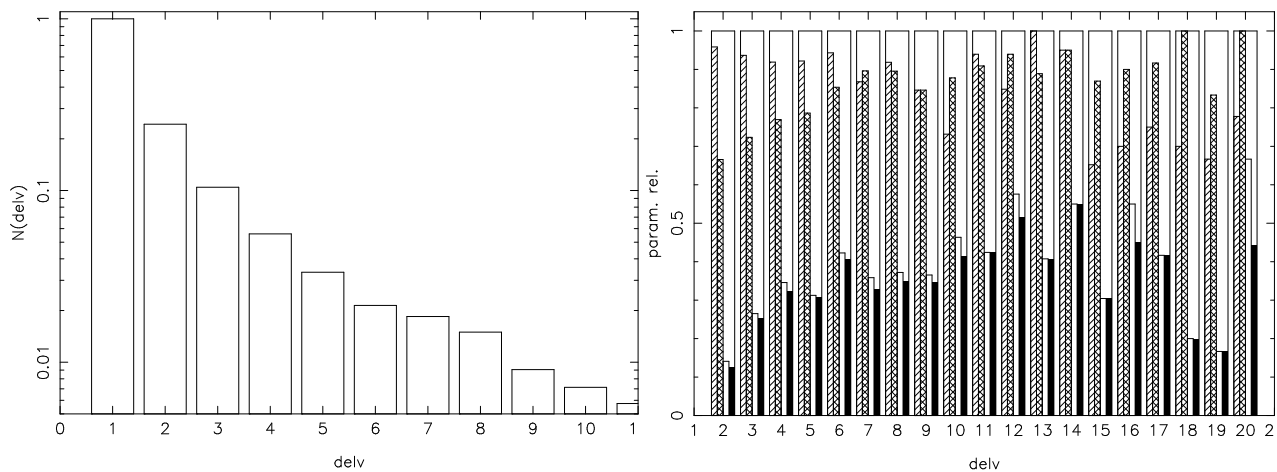

\centering
\includegraphics[angle=-90,width=0.45\linewidth]{shell_fig01a.eps} 
\includegraphics[angle=-90,width=0.45\linewidth]{shell_fig01b.eps}
    \caption{Relative number of structures identified in LAB as a function
             of their velocity extent (left panel); the relative number of 
             structures that fulfill the $P_{2,\mathrm{min}}$ criterion 
             (hatched),
             $P_{2,\mathrm{ave}}$ criterion (cross-hatched),  
             $P_3$ criterion (empty), and all three
             (solid) 
             as a function of the extent of velocity (right panel).}
       \label{ln1_hist1}
\end{figure*}

\subsubsection{Definition of a 2D structure}

In channel maps we search for regions with locally low brightness 
temperatures. The region has to be continuous, i.e. all pixels that belong
to it must
be connected by their edges, not only by touching their corners, and their
brightness temperatures must be lower than or equal to a certain level of
brightness temperature ${T_\mathrm{B}}_{\mathrm{level}}$. All neighbouring 
pixels, which are not members of 
the identified region, have to have brightness temperatures higher than 
${T_\mathrm{B}}_{\mathrm{level}}$. These conditions allow embedded clouds with 
a higher temperature.

For each channel map, the range of ${T_\mathrm{B}}_{\mathrm{level}}$ is examined, 
starting at a level
of noise and growing to the maximum brightness temperature in the map.
For the LAB processing we used the square root scale for the brightness
temperature (i.e. we worked with $\sqrt{T_\mathrm{B}}$, not directly with 
$T_\mathrm{B}$), the lowest ${T_\mathrm{B}}_{\mathrm{level}}$ we processed was 
0.64 K, $\sqrt{{T_\mathrm{B}}_{\mathrm{level}}}$ was 
increased by steps of 0.2.

The identified region has sizes in $l$ and $b$ directions. These sizes
have to be within limits (${d_{\mathrm{pix}}}_{\mathrm{min}}$,
${d_{\mathrm{pix}}}_{\mathrm{max}}$). For each region we find the highest 
${T_\mathrm{B}}_{\mathrm{level}}$ for which it is fulfilled. We used  
${d_{\mathrm{pix}}}_{\mathrm{min}} = 3$, which eliminated single
pixels and very small structures and sets the minimum size of a structure
to $1.5^{\degr}$. The value of ${d_{\mathrm{pix}}}_{\mathrm{max}}$, 
set to 90 (e.g. $45^{\degr}$), has no clear physical or numerical meaning, but 
we find out that it gives good results; we were able to identify a 
whole range of structures under many conditions occurring in the data. A 
lower value of ${d_{\mathrm{pix}}}_{\mathrm{max}}$ would prevent
identification of angularly large shells. Very few 2D structures have sizes 
larger than ${d_{\mathrm{pix}}}_{\mathrm{max}}=90$, only a few 
identifications in the solar neighbourhood, where such angularly large shells 
might occur.

\subsubsection{Creating a 3D structure}

For each 2D region identified in the previous step we check that it
has a counterpart in subsequent channel maps. We measure the overlap
between the neighbouring 2D structures, and if it is large enough,
we connect them in one 3D structure. This 
continues for all velocity channels and for all structures, until all
possible connections are made. 

The overlap condition is that at least a fraction $p$ of the
area of the smaller one of these two structures is overlapped. We used
a value $p=0.5$, but results were not very sensitive to this value
(we also tested values 0.1 and 0.9). If we use a low value of $p$,
we may artificially connect two or more physically unrelated, but neighbouring,
structures, while for a high value of $p$ we may divide one physical
structure with an irregular shape into several smaller pieces.

Connecting 2D structures into 3D ones may increase the angular sizes of 
structures, so they can exceed the value ${d_{\mathrm{pix}}}_{\mathrm{max}}$.

\subsubsection{Properties of HI shells}

We do not prescribe any desired shape for an HI shell, we only request
that structures, which we call shells, span a certain 
velocity extent and that their shapes do not change significantly between 
consecutive velocity channels.

The first condition is simple. We demand that the velocity extent 
of the structure $\Delta v$ is greater than a given value.
The velocity extent is
\begin{equation}
\Delta v = v_{\mathrm{max}} - v_{\mathrm{min}},
\end{equation}
where $v_{\mathrm{min}}$ and $v_{\mathrm{max}}$ delimit the interval in
which the structure exists.

To estimate how much the shape of the structure changes between
the velocity channels we calculate the parameter $P_2$ 
\begin{equation}
P_{2} (iv) = {{N_{\mathrm{both}}(iv,iv+1)} \over  
{{\mathrm{max}}(N_{\mathrm{pix}}(iv),N_{\mathrm{pix}}(iv+1))}},
\end{equation}
where $N_{\mathrm{pix}}(iv)$ is the number of pixels belonging to the structure 
in the channel
$iv$, $N_{\mathrm{both}}(iv,iv+1)$ is the number of overlapping pixels, i.e. 
those pixels with positions $il$, $ib$, which belong to the structure in both
channels $iv$ and $iv+1$. 

If $P_2$ has a low value, it means that the $lb$-size or the $lb$-position
in subsequent velocity channels are very different. It may also mean
that the 3D shape of the structure changes substantially
between the two neighbouring channels. 
It serves as a warning that the structures might not be compact or uniform.
Since an ideal expanding shell, viewed in velocity channels, starts small,
reaches its maximum size, and then shrinks again, we determine
the average $P_{2,\mathrm{ave}}$ and the minimum value $P_{2,\mathrm{min}}$, not 
in the whole velocity range of the structure, but only in the limited range 
of velocity channels between
the two local maxima in $P_2$, the first after the lowest velocity channel,
the second the last before the highest velocity channel. 

The calculation of $P_2$ is similar, but not identical, to the calculation
of the overlapping parameter $p$ in the previous step. The difference
is that 1) here we compare the number of overlapping pixels with the larger 
of these
two channel structures (it was the smaller one previously), and 2) in this 
step we already evaluate a structure constructed from several originally 
individual 2D parts. 

The spectrum through the centre of the shell should be seen as a noticeable, 
smooth depression in the brightness temperature profile compared 
to the profile outside the shell. We construct the relative profile
\begin{equation}
\Delta T_{\mathrm{B}} (iv) = |T_{\mathrm{in}}(iv) - T_{\mathrm{out}}(iv)|,
\end{equation}
where $T_{\mathrm{in}}$ is the average brightness temperature in pixels
belonging to the structure in a given velocity channel and $T_{\mathrm{out}}$
is the average value of $T_\mathrm{B}$ in pixels just outside the structure.
Which pixels are outside is determined for each velocity channel separately
to account for the changing $lb$-size of the structure. Then we define
\begin{equation}
P_3 = \mathrm{max}_{v}(\Delta T_\mathrm{B}),
\end{equation}
i.e., the maximum difference in average values of brightness temperature
inside and outside the shell in all channels where the given structure exists.
Clearly visible shells have high $P_3$. Low $P_3$ is connected with poorly
visible shells, but
also with shells in a region with high $T_\mathrm{B}$ gradients. 

The 3D structure is claimed to be an HI shell if
it fulfills the following three conditions:
\begin{equation}
\Delta v \geq \Delta v_{\mathrm{min}}; \\
P_{2,\mathrm{min}} \geq  \mathcal{P}_{2,\mathrm{min}}; \\
P_{2,\mathrm{ave}} \geq  \mathcal{P}_{2,\mathrm{ave}}; \\
\end{equation}
with a possible auxiliary ``visibility'' parameter,
\begin{equation}
P_{3} \geq  \mathcal{P}_{3},
\end{equation}
where the values of $\Delta v_{\mathrm{min}}$, $\mathcal{P}_{2,\mathrm{min}}$, 
$\mathcal{P}_{2,\mathrm{ave}}$, and $\mathcal{P}_3$ determine the strictness of 
the search.

\subsection{A new version of the identifying algorithm}

The numerical code described in this paper is a successor to the code
described and used in Ehlerov\'a \& Palou\v{s}, 2005. The main differences
follow.
\begin{enumerate}
\item{The current version deals with the whole datacube at once, while the old
      version had to cut the big datacube into smaller subcubes and then
      connect them back together.

Cutting the big datacube into smaller ones led to artificial edge problems
(most important to distortions in the shape of identified structures
and the failure of some identifications).

      }
\item{The last step in the identification, ``Properties of HI shells'', was
      completely changed. The old version used the analysis of the 
      $T_\mathrm{B}(v)$
      spectrum through the central pixel of the structures, the current
      version used the geometrical parameters $P_{2,\mathrm{min}}$ and 
      $P_{2,\mathrm{ave}}$,
      and an auxiliary visibility parameter $P_3$. Both versions used
      the condition with the $\Delta v_{\mathrm{min}}$, which was set in the 
      previous
      version to $\Delta v_{\mathrm{min}}= 4\ \mathrm{kms}^{-1}$.

The spectrum analysis worked well for small structures in 
smooth surroundings. For larger structures and a more violent
medium, it was sensitive to the exact position of the studied spectrum
and to the definition of the background enhancing the possibility of
false identifications.
Geometrical parameters from the current version are more robust. 
      }
\end{enumerate}

\section{Results}

\begin{figure*}
\centering
\includegraphics[angle=0,width=0.45\linewidth]{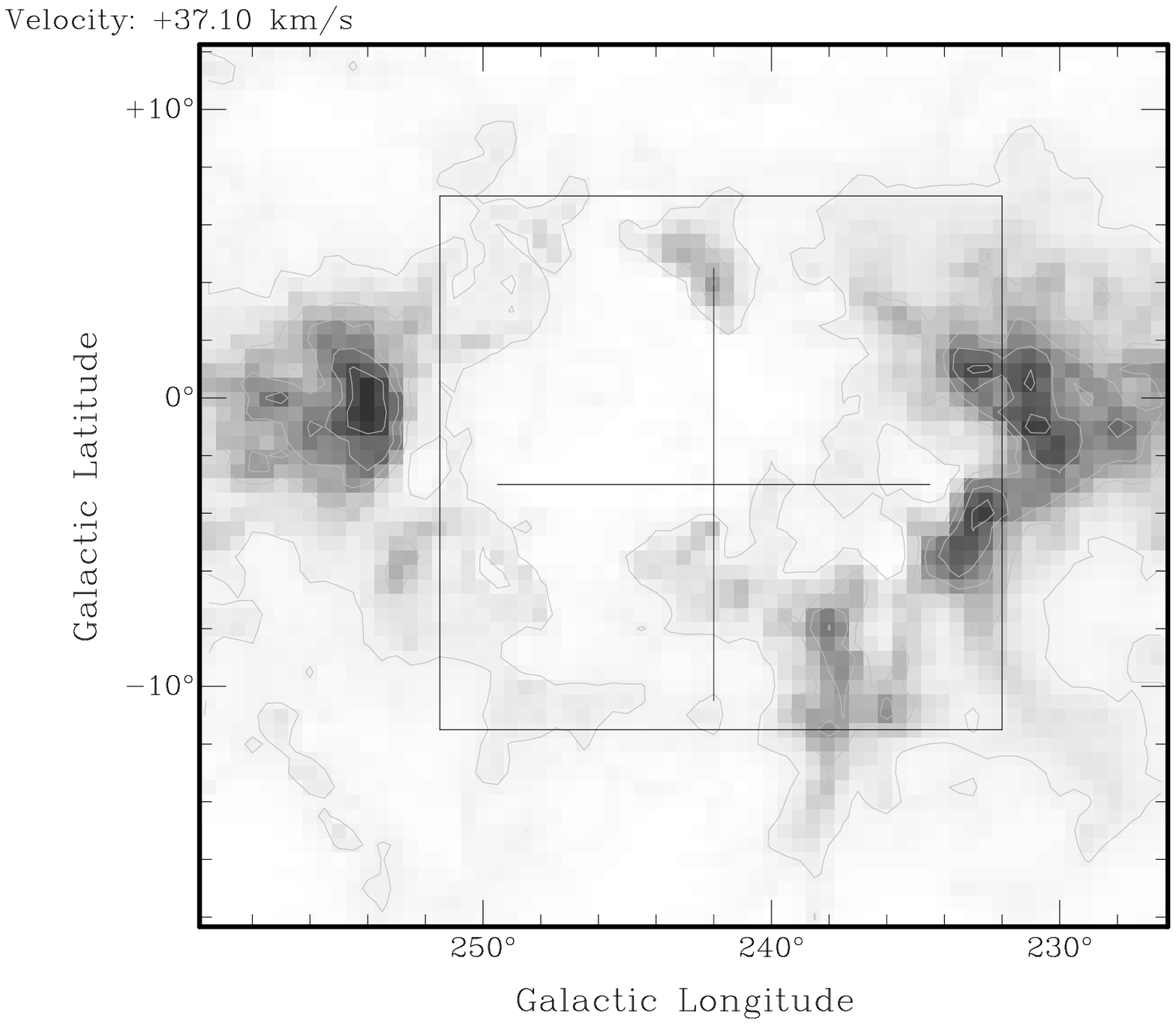} 
\includegraphics[angle=0,width=0.45\linewidth]{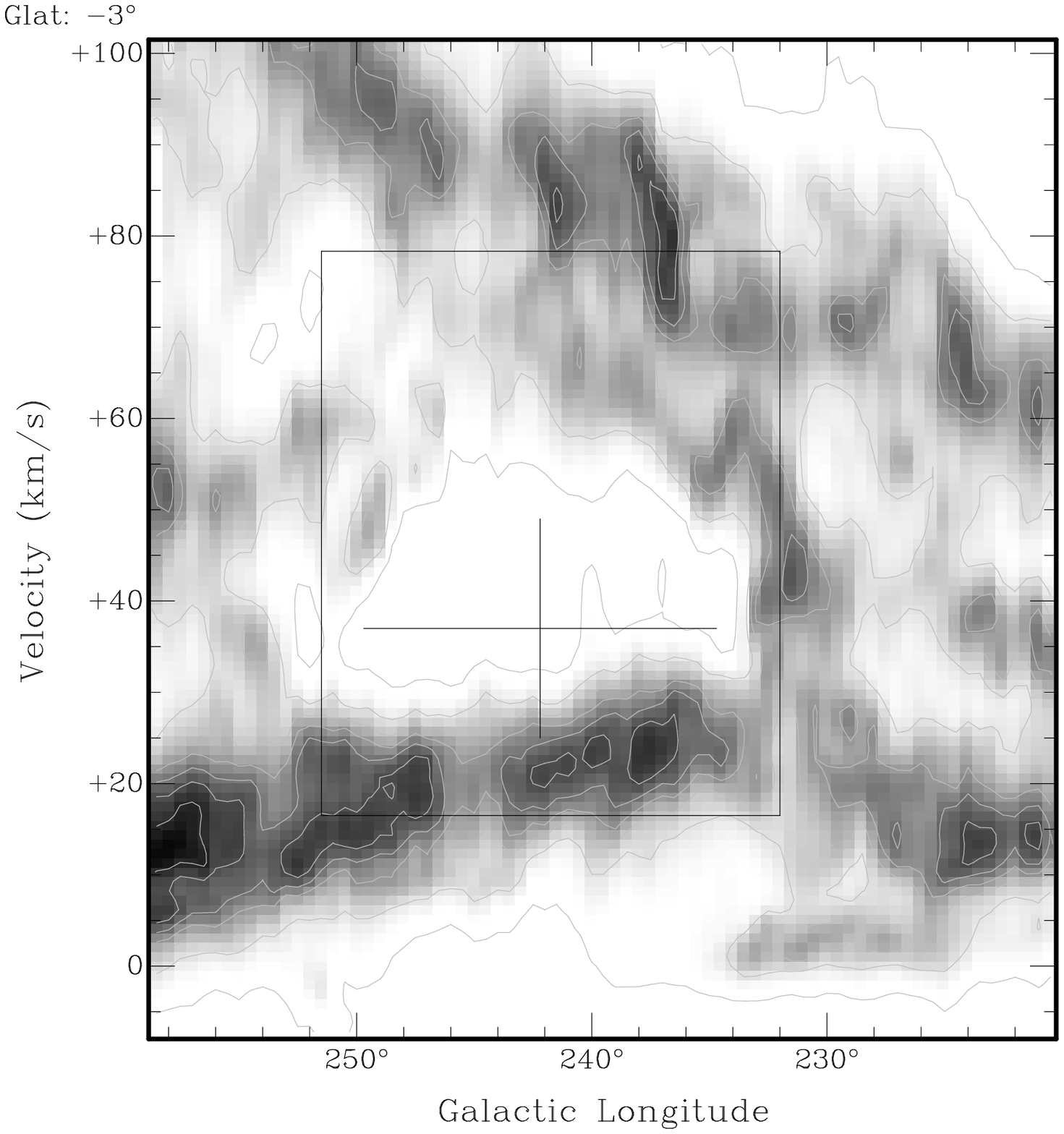}
    \caption{
     Region of GS242-02+37, one of the expanding shells in 
     \citet{1979ApJ...229..533H}, as seen in the LAB survey. The original 
     identification is shown by the black cross (length of its arms
     corresponds to published sizes),
     our new identification (the structure 
     GSH243.5-02.5+043.3 in appended tables) is shown
     by the box. Left panel is the $lb-$map, right panel is the $lv-$diagram.
    }
       \label{str02122}
\end{figure*} 

\begin{figure*}
\centering
\includegraphics[angle=0,width=0.45\linewidth]{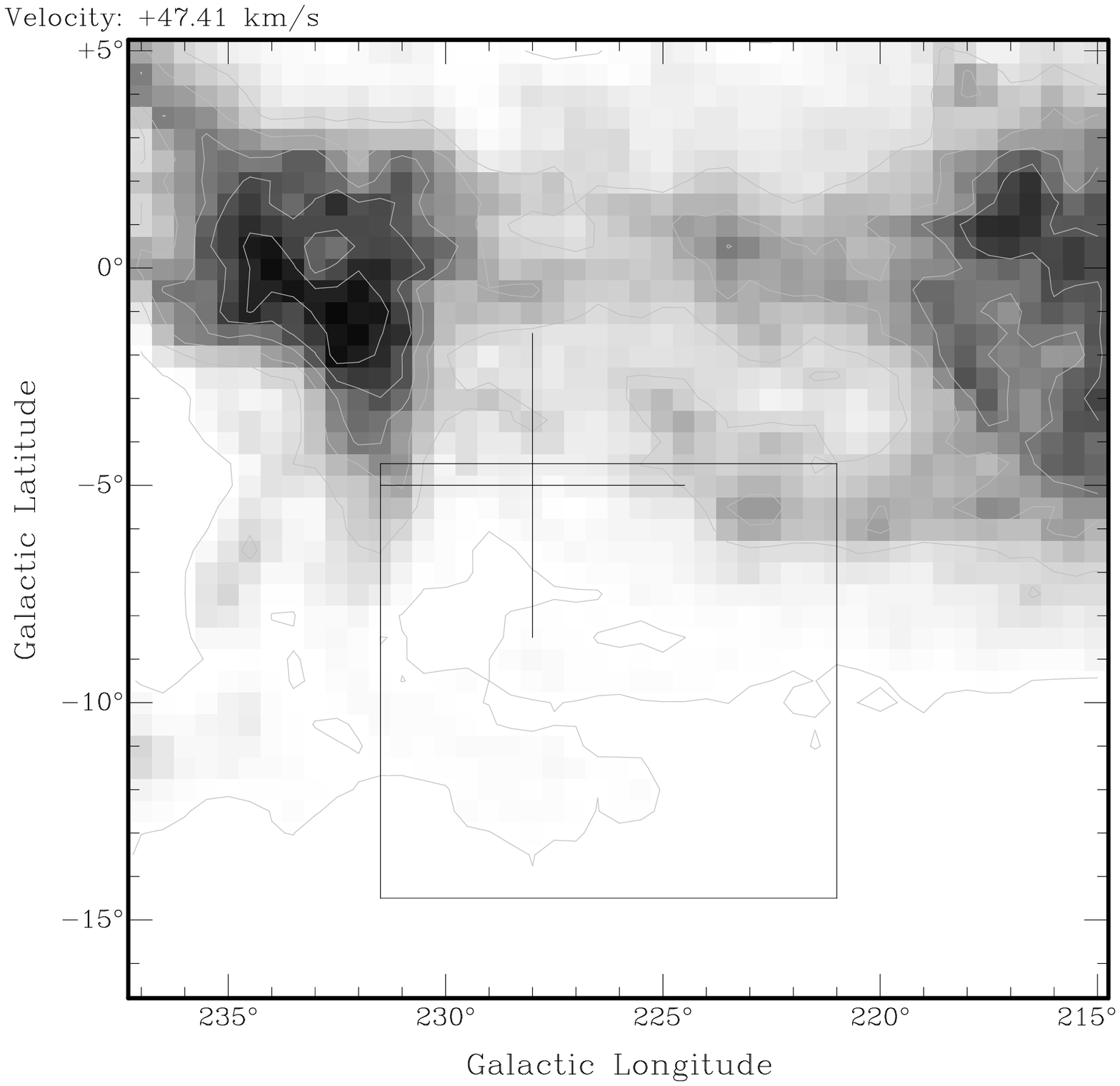} 
\includegraphics[angle=0,width=0.45\linewidth]{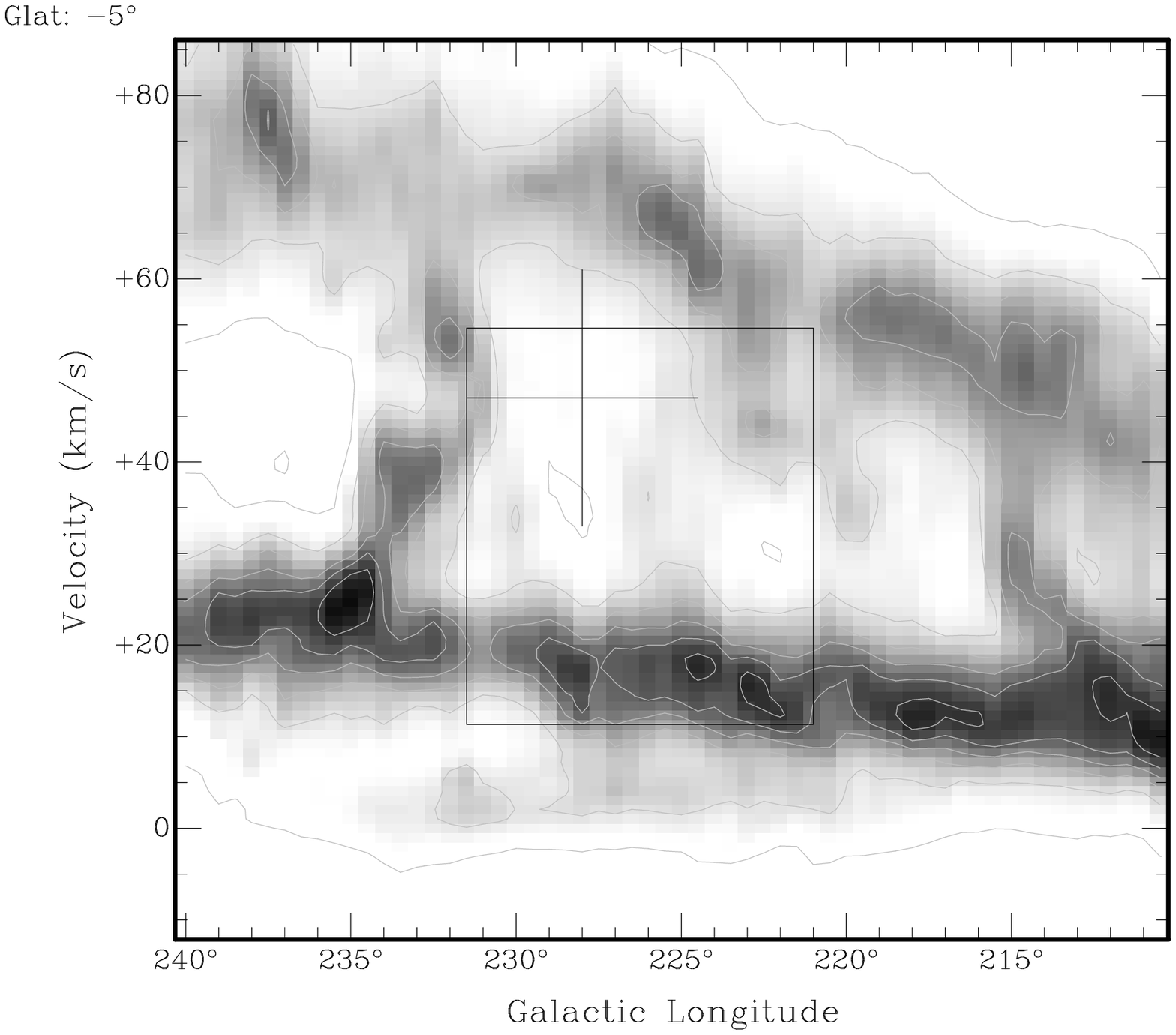}
    \caption{
     Region of GS228-05+47, one of the stationary shells in 
     \citet{1979ApJ...229..533H}, as seen in the LAB survey. The original 
     identification is shown by the black cross (length of its arms
     corresponds to published sizes), 
     our new identification (the structure 
     GSH226.5-09.5+031.9 in appended tables) is shown
     by the box. Left panel is the $lb-$map, right panel is the $lv-$diagram.
    }
       \label{str03121}
\end{figure*}

\subsection{LAB}

We used the LAB survey with 
$|v_{\mathrm{lsr}}| \le 250\ \mathrm{kms}^{-1}$ as an input
datacube for our identifying algorithm. Using 
${d_{\mathrm{pix}}}_{\mathrm{min}} = 3$ and
${d_{\mathrm{pix}}}_{\mathrm{max}} = 90$ as a minimum and a maximum size of the 
structures and the overlapping parameter $p=0.5$, we identified 17535 
3D structures. The majority of these structures span only a few velocity 
channels (see Fig. \ref{ln1_hist1}, left panel): 
2152 structures (12 \%) have a velocity extent greater than or equal to 
four channels (4.1 $kms^{-1}$), 775 structures (4.4 \%) have an extent greater 
than or equal to eight channels (8.2 $\mathrm{kms}^{-1}$). We eliminated 
irregular 
structures by adopting parameters $\mathcal{P}_{2,\mathrm{min}} = 0.2$ and 
$\mathcal{P}_{2,\mathrm{ave}} = 0.5$ (see Fig. \ref{ln1_hist1}, right panel,  
showing 
the fraction of structures, which fulfill the required criteria).
Geometrical conditions $\mathcal{P}_{2,\mathrm{min}}$ and 
$\mathcal{P}_{2,\mathrm{ave}}$ 
eliminate fewer than $\sim 40 \%$ of the shells (depends on their 
velocity extent). As shown in Fig. \ref{ln1_hist1}, 
the most restrictive parameter is the visibility parameter $\mathcal{P}_3$:
only about $40-50 \%$ of shells have parameter $P_3$ greater than
4 K.  The value of $\mathcal{P}_3 = 4\ K$ will be justified in the
next section.

The most important criterion in eliminating shells is the minimum required 
velocity extent $\Delta v_{\mathrm{min}}$. The lowest acceptable value is 
probably 
$\Delta v_{\mathrm{min}} = 4\ \mathrm{kms}^{-1}$, and the safe option is 
closer to 
$\Delta v_{\mathrm{min}} = 8\ \mathrm{kms}^{-1}$.
This parameter is related to the dispersion velocity in the ISM.
A value of $\Delta v_{\mathrm{min}}$ lower than the typical 
velocity dispersion leads to identification of some false structures,
while the higher value may eliminate some real structures. The dispersion
velocity in ISM varies greatly, depending on the conditions in ISM, but
its typical value for the HI emission is around $7\ \mathrm{kms}^{-1}$.

\subsection{Comparisons to previously known HI shells}

\begin{table*}
\caption{Comparison of identifications with HI supershells:
  the whole Galaxy (the middle section), the 2nd and 3rd quadrants (the
  right section).
}  
\label{tab_comp1}      
\centering                          
\begin{tabular}{|l|rrr|rrr|}        
\hline
 ~ & $num$       & vol. restr. & vol. unrestr. 
   & $num_{Q2+3}$ & vol. restr. & vol. unrestr.  \\
\hline
Heiles 1979, stationary & 
       46 & 14 (30\%) & 33 (72\%) & 18 & 6 (33\%) & 15 (83\%) \\
Heiles 1979, expanding  & 
       17 &  7 (41\%) &  9 (53\%) &  7 & 4 (57\%) &  4 (57\%) \\
McClure-Griffiths et al 2002 
               & 19 & 10 (53\%) & 17 (89\%) &  4 & 3 (75\%) &  4 (100\%) \\
\hline
\end{tabular}
\tablefoot{The left section denotes sources, from which previously
    known HI shells
    were taken. The middle and right sections give the total number of 
    shells (in the whole Galaxy or in the 2nd and 3rd quadrant, respectively)
    and the number of shells, which were identified again by our method  
    (percentage of the total is in parentheses). We distinguish two cases:
    the first,  where we demand that sizes of the compared
    structures are similar (vol. restr.), and the second, where we do not 
    put any restrictions on sizes of the structures (vol. unrestr.). }
\end{table*}

We compared our LAB identifications to two previous papers on HI shells
in the Milky Way: \citet{1979ApJ...229..533H} and \citet{2002ApJ...578..176M}, 
which contain the majority of known HI supershells. Heiles's list is divided 
into two parts: stationary shells, which do not change their size with 
velocity;
and expanding shells, which show different sizes in different velocity
channels.

Table \ref{tab_comp1} shows a comparison of
82 supershells from the two papers with our LAB identifications 
for the whole datacube (the middle section of the table) and separately 
for the second and third quadrants (the right section of the table).
We chose these two quadrants, because they contained  only the outer 
Galaxy, where structures are less crowded, which makes identifications
more accurate. We distinguish whether the shells have similar
sizes (``volume restricted'') or not (``volume unrestricted'').

The best agreement was for the McClure-Griffiths shells (in all cases
more than 50 \% of supershells were re-identified), the worst agreement for
Heiles' stationary shells. This is due to the quality of
data used in the different searches. Stationary shells were re-identified with 
different sizes more frequently than expanding shells: 
the trace of a stationary shell in a datacube
is less sharply defined, less pronounced, and not as specific as the trace of an
expanding shell, and therefore it is easily mistaken by both human eyes and
numerical methods. Figures \ref{str02122} and \ref{str03121} show one expanding
and one stationary shell from the list by \citet{1979ApJ...229..533H} and 
the corresponding LAB 
identifications. Comparisons in the outer Galaxy were more
favourable. This is, we believe, because the ISM in the inner Galaxy is
very violent, and there is an overlap from near and far regions along the
line of sight, leading to problems in artificial connections
of unrelated structures, all of which make identifications in the inner Galaxy
less reliable.

We conclude that our method is reliable in the outer Milky Way, because we 
are able to identify the majority of known expanding HI supershells
and a part of known stationary HI supershells again. However, even in the outer
Milky Way
the existence of some of these published shells is doubtful.
Our method is less reliable in the inner Milky Way. 

All LAB identifications connected to the previously known HI supershells
had $\Delta v \geq 9\ \mathrm{kms}^{-1}$, $P_{2,\mathrm{min}}\geq 0.2$, 
$P_{2,\mathrm{ave}}\geq 0.7$, and $P_3 \geq 4.0\ K$. 

\begin{figure}
 \centering
 \includegraphics[angle=-90,width=0.45\textwidth]{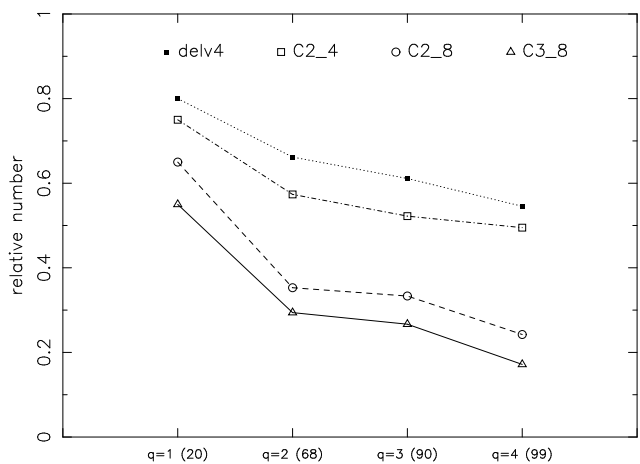}
 \caption{Comparisons between previous and current versions of the search
          algorithms. Different symbols
          denote different sets of parameters put on new
          identifications (see the text for the exact explanation of 
          delv4, C2$\_{}$4, C2$\_{}$8, and C3$\_{}$8).
          The $y$-value shows what fraction of old identifications is 
          also observed with the new algorithm.
         }
 \label{holmes_oldnew}
\end{figure}

\begin{figure*}
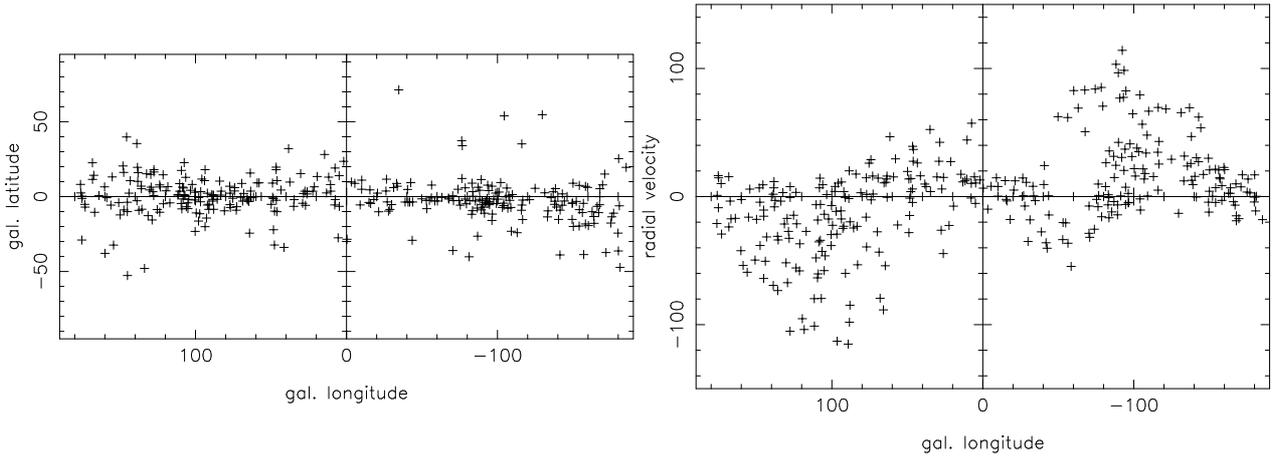

\centering
\includegraphics[angle=-90,width=0.45\linewidth]{shell_fig05a.eps} 
\includegraphics[angle=-90,width=0.45\linewidth]{shell_fig05b.eps}
    \caption{Distribution of 333 identified shells on the sky. Left panel:
    lb-map; right panel: lv-map.}
       \label{lb_lv_c3_8}
\end{figure*}

\subsection{Comparison to the previous version of the search algorithm}

The previous version of the search algorithm was applied to the
Leiden-Dwingeloo HI survey 
\citep[only the northern sky,][]{2005A&A...437..101E}. We 
compared current LAB identifications with our previous
identifications. We restricted ourselves to the second quadrant, which 
was fully observed in the previous version and which did not suffer from
shell crowding in the inner Galaxy.

We compared what fraction of shells identified by the previous method
(277 in total in the 2nd quadrant) was re-identified by the new method
with various values of search parameters
$\Delta v_{\mathrm{min}}$, $P_{2,\mathrm{ave}}$ (the average overlap), 
$P_{2,\mathrm{min}}$ (the minimum overlap), and $P_3$ (visibility parameter). In
Fig. \ref{holmes_oldnew} we show the comparison for
\begin{enumerate}
\item{$\Delta v_{\mathrm{min}} = 4\ \mathrm{kms}^{-1}$ 
   (designated as delv4 in Fig. \ref{holmes_oldnew}, an undemanding
   condition);}
\item{$\Delta v_{\mathrm{min}} = 4\ \mathrm{kms}^{-1}$, 
   $P_{2,\mathrm{min}} \geq 0.2$, $P_{2,\mathrm{ave}} \geq 0.5$
   (designated C2$\_{}$4 in Fig. \ref{holmes_oldnew});}
\item{$\Delta v_{\mathrm{min}} = 8\ \mathrm{kms}^{-1}$, 
   $P_{2,\mathrm{min}} \geq 0.2$, $P_{2,\mathrm{ave}} \geq 0.5$
   (designated as C2$\_{}$8 in Fig. \ref{holmes_oldnew});}
\item{$\Delta v_{\mathrm{min}} = 8\ \mathrm{kms}^{-1}$, 
   $P_{2,\mathrm{min}} \geq 0.2$, $P_{2,\mathrm{ave}} \geq 0.5$,
   $P_3 = 4\ \mathrm{K}$ 
   (designated as C3$\_{}$8 in Fig. \ref{holmes_oldnew}, the
   strictest set of conditions).} 
\end{enumerate}

Furthermore, we demanded that sizes of shells were similar (``volume
restricted'').
Corresponding fractions were 61 \% (the 1st set of conditions),
54 \% (2nd), 33 \% (3rd), and 26 \% (4th). Each structure in the previous
version had a visual quality, which describes how good it looked  (a purely
subjective criterion:
a shell with $q=1$ is clearly visible in velocity channels and
is probably spherical or elliptical; a shell with $q=4$ is inconspicuous
in velocity channels, with a low contrast to its surrounding and probably
fragmented and irregular).
The comparison was performed separately for shells
with different qualities (Fig. \ref{holmes_oldnew}), and the best-looking 
shells (visual quality 1) had corresponding fractions ranging from 
80 \% to 55 \%.
Fractions of reidentified shells for volume-unrestricted comparisons were
88 \% (the 1st set of conditions), 81 \% (2nd), 76 \% (3rd), and 72 \% (4th).

We also compared what fraction of those shells that fulfilled the given
set of conditions were also classified as shells by the previous method
(Fig. \ref{holmes_oldnew}). The fractions are 25 \% for the first set of 
conditions, 29 \% (2nd), 48 \% (3rd), and 74 \% (4th). That means that
three fourths of shells fulfilling the strictest set of conditions were also 
defined as shells by the previous method. One fourth are new: these
shells were rejected because of edge effects inherent to the previous 
version of the code (described in the Sect. 2.3).

The summary of these results follows. The less strict the set of conditions
used for the new identifications, the better the agreement we get with
the previous version of the searching algorithm. Visually nice shells
are much less sensitive to the precise choice of conditions and methods.
Differences arise for the less pronounced and not very regular structures. 

By using the fourth set of conditions 
($\Delta v_{\mathrm{min}} = 8\ \mathrm{kms}^{-1}$, 
$P_{2,\mathrm{min}} \geq 0.2$, $P_{2,\mathrm{ave}} \geq 0.5$, 
$P_3 = 4\ \mathrm{K}$), we chose
structures that have regular, not wildly changing shapes
that are noticeable and have
a large velocity extent. All previously known supershells
belong to this group, and the majority of shells from this group were also 
identified
by the previous method. This is the default setting for any further discussion.

\section{Discussion}

\begin{figure*}
 \centering
 \includegraphics[angle=-90,width=\textwidth]{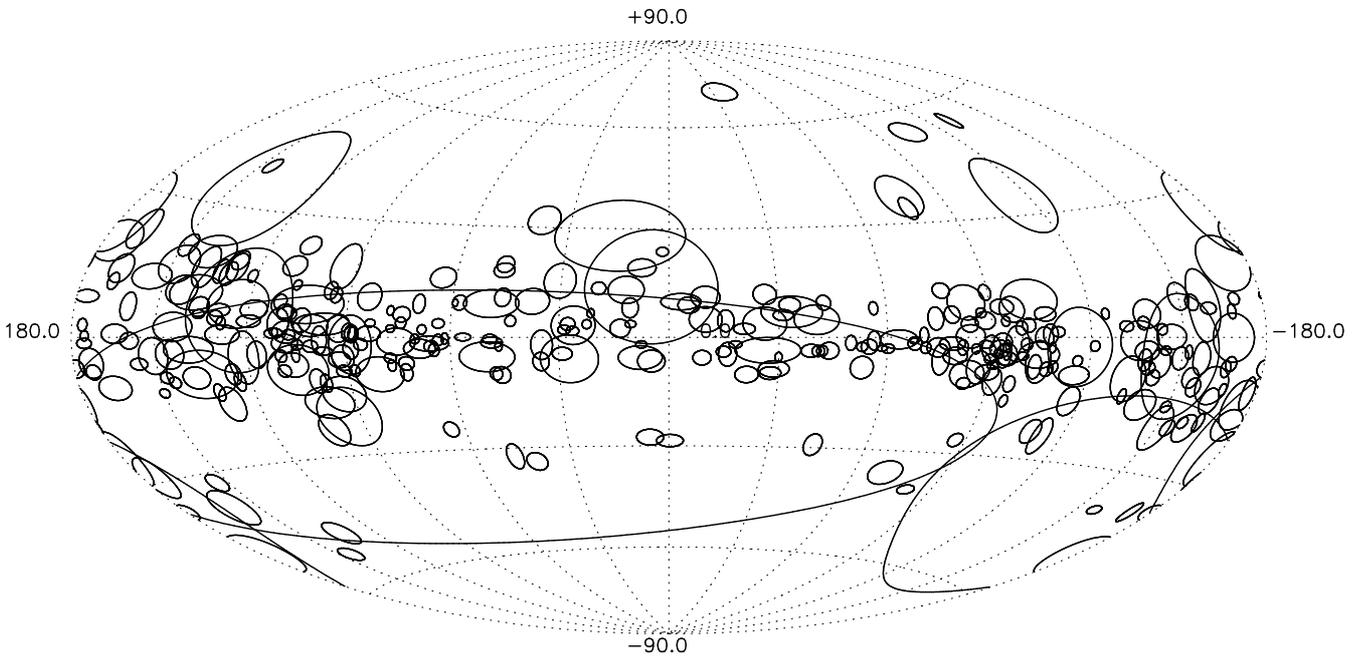}
 \caption{Identified HI shells, Hammer-Aitoff projection, centered
          on $l=0^{\degr}$, $b=0^{\degr}$.}
 \label{sky1}
\end{figure*}

\begin{figure}
 \centering
 \includegraphics[angle=-90,width=0.45\textwidth]{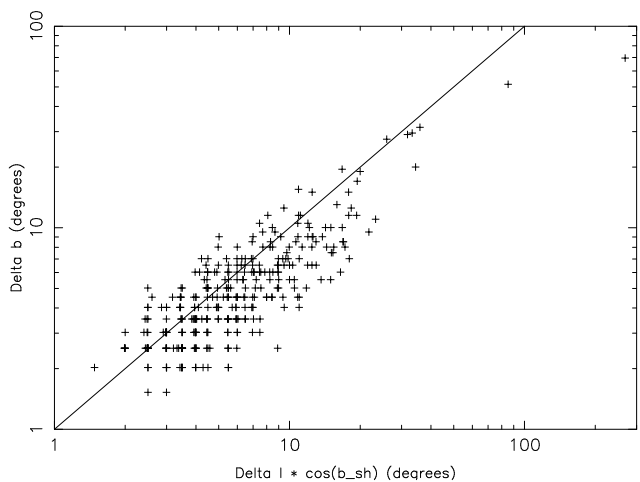}
 \caption{Angular sizes of identified shells.}
 \label{dldb}
\end{figure}

\subsection{Sky distribution of identified shells}

We identify 333 shells in the LAB data, which fulfill the following
criteria : $\Delta v_{\mathrm{min}} = 8\ \mathrm{kms}^{-1}$ 
(the minimum velocity span); 
$P_{2,\mathrm{min}} \geq 0.2$, $P_{2,\mathrm{ave}} \geq 0.5$ (geometrical 
parameters); and $P_3 = 4\ \mathrm{K}$ (the visibility parameter). Their 
positions on the
sky and in velocity space are shown in Figs. \ref{lb_lv_c3_8} and \ref{sky1}.
Ellipses in Fig. \ref{sky1} correspond to sizes of shells. 
Angular sizes of shells are shown in Fig. \ref{dldb}. The majority of 
structures are elongated in the $l$ direction rather than in $b$,
which may be explained as a consequence of the lower shell column
densities in high $b$ directions.

Online Table \ref{table1_onl}, available at the CDS, contains the names 
and positions of all identified structures: the name with coordinates of 
the centre; and the minimum and maximum $l$, $b$ and $v_{\mathrm{lsr}}$, 
where the structure is visible. 

\subsection{Galactic distribution of shells}

\begin{figure}
\centering
\includegraphics[angle=-90,width=0.9\linewidth]{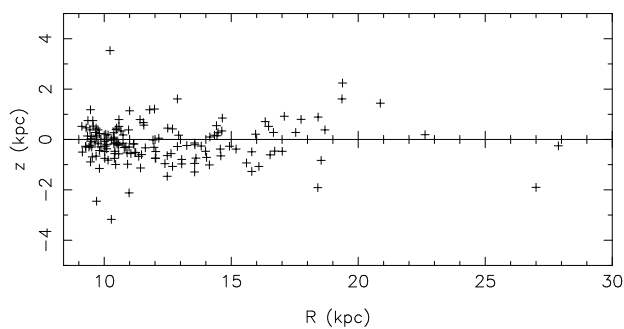}
    \caption{Positions of identified supershells: the galactocentric
     distance $R$ and the distance from the Galactic plane $z$.}
    \label{gal2_rz}
\end{figure}

\begin{figure}
\centering
\includegraphics[angle=-90,width=0.9\linewidth]{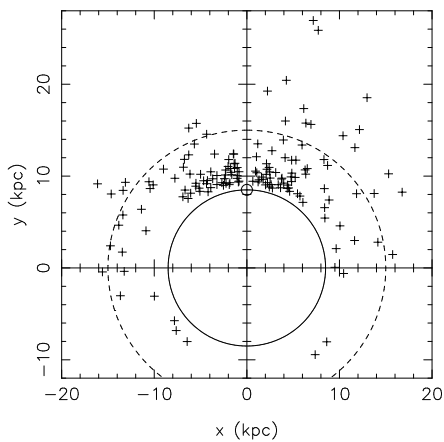} 
    \caption{Distribution of supershells in the Galactic plane. The position
     of the Sun is shown and two circles with galactocentric distances
     of 8.5 kpc and 15 kpc are overlaid.}
       \label{gal2_xy}
\end{figure}

The kinematic distances of identified HI shells are derived using
the rotation curve of \citet{1993A&A...275...67B}, assuming 
$R_{\sun} = 8.5\ \mathrm{kpc}$,
$V_{\sun}= 220\ \mathrm{kms}^{-1}$. 
We consider only shells in the outer Galaxy.
We exclude shells found at very low velocities 
$|v_{\mathrm{lsr}}| < 10\ \mathrm{kms}^{-1}$, since identifications at these 
low velocities
are heavily influenced by the local emission. We also disregard
shells close to the centre and anticentre directions
($l < 20^{\degr}$, $l > 340^{\degr}$, $170^{\degr} < l < 190^{\degr}$)
and shells with very high $b$ coordinates ($|b| > 50^{\degr}$), since 
noncircular motions may be comparable to projected rotational motions.
These selection criteria concern the following analyses and figures
from Fig. \ref{gal2_rz} onwards only.

Online Table \ref{table2_onl}, available at the CDS, contains the 
following information
about these shells: the galactocentric distance $R$, the radius of the
shell $r_{\mathrm{sh}}$ (calculated as a geometrical mean of sizes
in $l$ and $b$ directions), the expansion velocity $v_{\mathrm{exp}}$ 
(calculated as half of the measured velocity extent), the estimated
energy input $L$ needed to create the shell based on the solution
of \citet{1977ApJ...218..377W}, assuming the density from 
\citet{2008A&A...487..951K},
and the approximate age of the shell $t_{\mathrm{exp}}$ \citep[based on the
solution of][]{1977ApJ...218..377W}.

Figures \ref{gal2_rz} and \ref{gal2_xy} show galactic coordinates of
shells and their positions in the galactic plane. At galactocentric
distances greater than 19 kpc, there are shells only in the second quadrant,
others are nearly empty. This is especially striking for the third quadrant,
which is, from the geometrical point of view, equivalent to the second one. 

\begin{figure}  
\centering
\includegraphics[angle=-90,width=\linewidth]{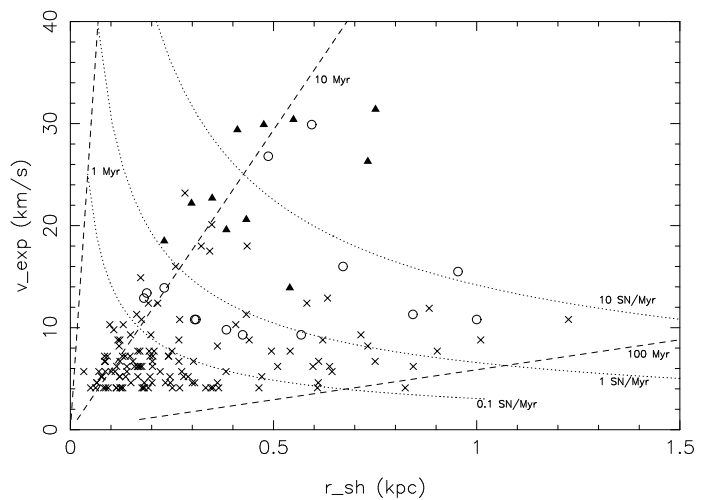} 
    \caption{$rv$ diagram of supershells. Evolutionary lines 
     \citep{1977ApJ...218..377W}
     for three luminosities are overlaid (dotted lines), as well as 
     three lines of different times (dashed lines), for the density 
     $n_0 = 0.3\ \mathrm{cm}^{-3}$. Different symbols denote estimated 
     energy inputs to shells, when real densities are taken into account
     (see text for the explanation): 
     crosses for $L < 0.1\ \mathrm{SNMyr}^{-1}$,
     open circles for $0.1\ \mathrm{SNMyr}^{-1} \leq L \leq 
     1.0\ \mathrm{SNMyr}^{-1}$,
     triangles for $L > 1.0\ \mathrm{SNMyr}^{-1}$.
     }
       \label{gal2_rv}
\end{figure}

Figure \ref{gal2_rv} shows the $rv$ diagram, i.e. the
dependence of the expansion velocity of the shell on its radius.
The expansion velocity of the shell is calculated as $0.5\Delta v$, and the 
radius is the geometric mean of sizes in $l$ and $b$ directions.
Using the analytical solution for the shell evolution 
\citep{1977ApJ...218..377W},
we derive the required energy inputs for individual shells and their ages, 
using the density  based on the distribution
of \citet{2008A&A...487..951K} and the warp parameters of 
\citet{2006ApJ...643..881L}. 
We find that 11 shells require energy inputs 
higher than $\sim 1\ \mathrm{SNMyr}^{-1}$. Calculated energy inputs are
given by different symbols in Fig. \ref{gal2_rv}. 

The analytical solution 
\citep[][lines in Fig. \ref{gal2_rv}]{1977ApJ...218..377W}
is overlaid for three different luminosities
($0.1\ \mathrm{SNMyr}^{-1}$,$1.0\ \mathrm{SNMyr}^{-1}$,
      $10.0\ \mathrm{SNMyr}^{-1}$),
assuming the ISM density $n_0 = 0.3\ \mathrm{cm}^{-3}$ (which is an 
overestimate for the majority
of shells) and for three different evolutionary times
($1\ \mathrm{Myr}$, $10\ \mathrm{Myr}$, and $100\ \mathrm{Myr}$). This
solution serves as a guideline, especially for ages, since the evolutionary
age in \citet{1977ApJ...218..377W} does not depend on density and luminosity,
while the calculated luminosity depends linearly on the density. 
We detect a few shells with ages below 5 Myr,
which is given by the resolution of the survey and constraints on the
sizes. There is only one shell with age greater than 100 Myr.
This lack of old shells may be explained by dissolution due to 
random motions and turbulence or due to destruction by spiral arms, 
differential rotation, and other galactic forces.

\subsubsection{Size distribution of shells}

\begin{figure}
\centering
\includegraphics[angle=-90,width=0.9\linewidth]{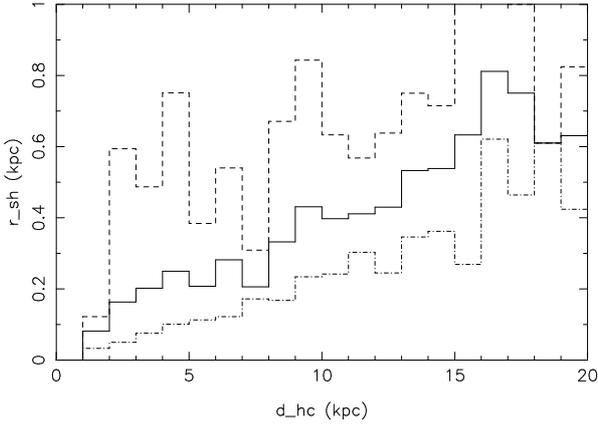} 
    \caption{Dependence of minimum (dash-dotted line), 
    average (solid line), 
    and maximum (dashed line)
    radius of the shell $r_{\mathrm{sh}}$ as a function of the heliocentric
    distance $d_{\mathrm{hc}}$.
    }
       \label{dhc}
\end{figure} 

We assume a power-law distribution for sizes of shells
\begin{equation}
\frac{dN(r_{\mathrm{sh}})}{dr_{\mathrm{sh}}}
  = \mathcal{A} {r_{\mathrm{sh}}}^{-\alpha}.
\end{equation}
We can then calculate the average size of shells 
$\overline{r_{\mathrm{sh}}} (d_{\mathrm{hc}})$
at a given heliocentric distance $d_{\mathrm{hc}}$ using observed
minimum and maximum sizes $r_{\mathrm{min}}(d_{\mathrm{hc}})$ and 
$r_{\mathrm{max}}(d_{\mathrm{hc}})$:
\begin{eqnarray}
\overline{r_{\mathrm{sh}}} (d_{\mathrm{hc}}) &=&
\frac{\int_{r_{\mathrm{min}}}^{r_{\mathrm{max}}} 
 N(r_{\mathrm{sh}})r_{\mathrm{sh}} dr_{\mathrm{sh}}}
 {\int_{r_{\mathrm{min}}}^{r_{\mathrm{max}}} N(r_{\mathrm{sh}}) dr_{\mathrm{sh}}} 
 \nonumber \\
&=& \frac{\alpha-1}{\alpha-2}
r_{\mathrm{min}}(d_{\mathrm{hc}})
\frac{1-(\frac{r_{min}(d_{\mathrm{hc}})}{r_{max}(d_{\mathrm{hc}})})^{\alpha-2}}
{1-(\frac{r_{\mathrm{min}}(d_{\mathrm{hc}})}
{r_{\mathrm{max}}(d_{\mathrm{hc}})})^{\alpha-1}}.
\end{eqnarray}

Since we restrict the minimum and maximum angular size on identified
shells, $r_{\mathrm{min}}$ and $r_{\mathrm{max}}$ must depend
on the distance (see Fig. \ref{dhc}, left panel). 
With known or assumed $\overline{r_{\mathrm{sh}}}$, $r_{\mathrm{min}}$, and 
$r_{\mathrm{max}}$ (assumptions are connected to restrictions on angular
sizes of detected structures), we can calculate $\alpha$. 
Solutions lie in the interval (2.2,3.1), the best solution is for 2.6.
This value might be artificially flatter than the real value, since
our identifying algorithm forms one big shell from two small shells
more frequently than vice versa, producing a flatter size spectrum.

An alternative method to calculate the coefficient $\alpha$, used
in Paper 1, is to take such a subset of identified shells, which in
a certain range of distances contains  shells in a given range of 
sizes. Results of this method cover an interval of $\alpha$ and
include the value 2.6 calculated here. Uncertainties in the 
determination of $\alpha$ are fairly large, probably because our
identifying algorithm is better at localizing structures than at
determining their precise sizes.

The average value of $\alpha $ in external galaxies 
\citep{2011AJ....141...23B}
is 2.9, which is a slightly steeper value than ours. Their value is an
average through several types of galaxies and there is a dependence of the
coefficient $\alpha $ on the galaxy type (early type spirals tend to have
steeper slopes). According to their Fig. 28, our value
of 2.6 corresponds to Sc galaxies.

The coefficient $\alpha$ is connected to the power-law coefficient $\beta$ 
of the luminosity function of sources creating shells, e.g. OB associations
\begin{equation}
\Phi(L) \propto  {L}^{-\beta}.
\end{equation}

According to \citet{1997MNRAS.289..570O} $\alpha$ and $\beta$ are connected as
\begin{equation}
\alpha = 2\beta -1.
\end{equation}
Therefore, $\beta \sim 1.8$, which is somewhat flatter
than the expected luminosity function of OB associations in the Milky Way 
\citep[2.0; see e.g.][]{1997ApJ...476..144M}.

\subsubsection{Radial distribution of shells}

\begin{figure}
\centering
\includegraphics[angle=-90,width=0.9\linewidth]{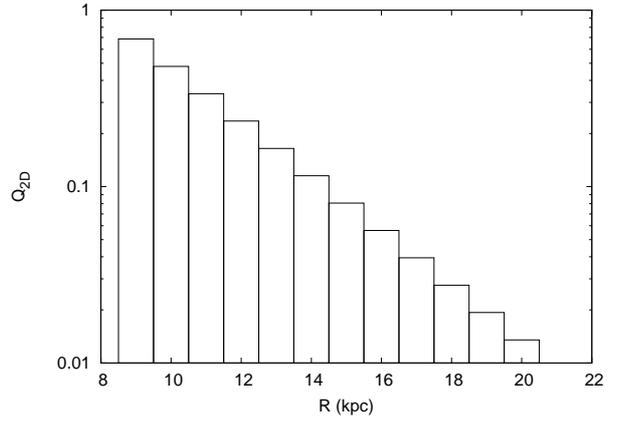} 
    \caption{2D porosity of the outer Milky Way.
     } 
    \label{porosity}
\end{figure}

We assume that the shell surface density follows an exponential distribution
\begin{equation}
\Sigma (R) = \Sigma_0 e^{-(R-R_{\sun})/\sigma_{\mathrm{gsh}}} 
\label{sigma_gsh}
\end{equation}
where $\Sigma (R)$ is the surface density of HI shells. 

Typical sizes of shells, hence the relative proportion
of identified to total number of shells, change with heliocentric distance,
and therefore to get intrinsic values of $\Sigma (R)$ we have to 
correct the observed values. With the corrected value we can calculate 
$\sigma_{\mathrm{gsh}}$. We apply the same method as in Paper 1, briefly 
described
in the Sect. 4.2.1. It is based on selecting subsets that
contain information for a chosen interval of heliocentric 
distances and sizes. The radial distribution of the shell surface density
$\Sigma (R)$ is based on
the number of shells with the galactocentric distance $R$
corrected for the limitations caused by a finite range of heliocentric 
distances. Results depend on the selected interval of distances, but
in a particular way: results, which are consistent 
with the assumed exponential decrease (i.e. show a high correlation
between observed numbers of shells with numbers predicted from
Eq. \ref{sigma_gsh})
form a distinct group, while the others form another one with a wide
spread in values of $\sigma_{\mathrm{gsh}}$. Good solutions
for $\sigma_{\mathrm{gsh}}$ 
have the average value of $(2.8 \pm 0.5)$ kpc.

After extrapolating solutions for the size distribution of shells
and the radial dependence of the shell surface density, we can estimate 
the total number of shells and also the porosity of the Galaxy. 
These extrapolations are not too precise, since they combine all
inaccuracies coming from errors in the derivation of $\alpha$ and 
$\sigma_{\mathrm{gsh}}$, and they also include the effect of selection
criteria described in the Sect. 3.
Taking the minimum size of a shell to be 100 pc and the maximum size 
to be 1 kpc, we come
to the conclusion that there should be something like 
400 to 1000 shells in the
outer Galaxy, and about 2 000 to 5 000 such shells in the whole Galaxy. 

The number surface density of HI shells with radii 
$r_{\mathrm{sh}}\geq 100\ pc$ in the solar neighbourhood is 
about $4^{+1}_{-2}$ shells per $\mathrm{kpc}^2$. 
Apart from uncertainties connected with extrapolating, there are
intrinsic variations in the surface density (see e.g. Fig.
\ref{gal2_xy}), most probably connected to spiral arms. The derived
value of surface density
is higher than --- though not incomparable to --- the
average value of the same quantity for a number of external galaxies 
\citep[THINGS galaxies,][]{2011AJ....141...23B}. The typical value is 
(0.1-1) shells per $\mathrm{kpc}^2$,
though there are some galaxies in that paper that reach higher values.
This discrepancy may be caused by either the less strict conditions
put on the identification method for
HI shells on our side and stricter conditions on theirs (among others, 
\citet{2011AJ....141...23B} demanded quite a high temperature contrast for 
their shells and regular elliptical shapes), or 
by a different point of view: from inside in the Milky Way and from outside
for external galaxies. The inside view is much more sensitive to
shells that did not yet break through the HI disc. On the other hand, the 
surface density of OB associations
in the solar neighbourhood is about 7 per $\mathrm{kpc}^2$ 
\citep{1997ApJ...476..144M}, 
which is comparable to our value: we would instead expect a higher number 
of shells than OB
associations, since the typical lifetime of a shell is longer than for
an OB association, but clustering of OB associations may decrease
the number of OB associations per shell.  

The porosity $Q$, which is a ratio of the surface occupied by shells to the 
total surface (the 2D porosity) or the volume occupied by shells to the total
volume (the 3D porosity), is derived. In the inner Galaxy the porosity as 
calculated from derived distributions, is large, much larger than 1, 
which predicts the significant overlapping of shells.
The porosity of the outer Galaxy is smaller (see Fig. \ref{porosity}),
where values lower than $\sim 0.01$ are reached at galactocentric distances 
of $\sim 20\ \mathrm{kpc}$.
Local values at the solar neighbourhood are $Q_{\mathrm{2D}} = 0.7$ and 
$Q_{\mathrm{3D}} = 0.4$
(assuming a cylinder with height 1 kpc) for shells with radii
$r_{\mathrm{sh}} \in (0.1,1.0)\ \mathrm{kpc}$. 
Error boxes are similar to those in the case of number surface density.
Porosity (especially $Q_{\mathrm{3D}}$) is mostly dependent on the maximum 
size of
the shell, which we assume to be 1 kpc, and therefore values quoted here
approach the real values, since the contribution of small bubbles is 
negligible.

\citet{2007ApJ...661..285D} calculate the partial porosity of shells with 
radii $r_{\mathrm{sh}} \in (5,40)\ \mathrm{pc}$ in the region of the Perseus arm
${Q_{\mathrm{3D}} = 0.007}^{+0.025}_{-0.003}$. We
can do the same thing by extrapolating our solutions to small shells
and get the result $Q_{\mathrm{3D}} = 0.003$. This is surprisingly good
agreement. If this is not a coincidence, it means
that we both detect the same type of objects - in different evolutionary
phases - even though we use substantially different approaches.  
Our somewhat lower value could mean that some of the small objects
will not evolve into large structures (such as bubbles created by single
supernova explosions).

\section{Conclusions}

We have identified 333 HI shells  
in the Leiden/Argentina/Bonn HI survey  of the Milky Way using
an automatic searching
algorithm. The advantage of this approach is the uniformity of 
conditions throughout the Milky Way. Our method is quite sensitive for locating
``something interesting'', while it is not as effective for measuring 
the precise sizes of the shells. 

We discovered the asymmetry between the second and third Galactic quadrants in 
the distribution of shells at large galactocentric distances.
At distances greater than 19 kpc there are shells only in the second quadrant.
The third quadrant, which is an equivalent to the second one from the 
geometrical point of view, contains no shells at these galactocentric distances.
 
Because of crowding and distance ambiguity in the Milky Way, we base the 
following
conclusions on shells in the outer Galaxy. We found that their radial 
distribution decreases exponentially on a scale length of 
2.8 $\mathrm{kpc}$, 
and their size distribution is a power law with the coefficient of $\sim 2.6$
(corresponding to a power-law coefficient of the luminosity function
of OB associations $\sim 1.8$).
The surface density of shells with radii $<0.1,1.0> \mathrm{kpc}$ in the solar
neighbourhood is $\sim 4\ \mathrm{kpc}^{-2}$.

\begin{acknowledgements}
This study has been supported by the Czech Science
Foundation grant 209/12/1795 and by the project RVO: 67985815.
This research made use of NASA's Astrophysics Data System.
The authors would like to thank the anonymous referee for helpful proposals
and Jim Dale and the A\&A language editor Joli Adams
for the help with the manuscript (any remaining inelegance
is purely our own).
\end{acknowledgements}

\bibliographystyle{aa} 
\bibliography{shell_ref}

\begin{thebibliography}{21}
\expandafter\ifx\csname natexlab\endcsname\relax\def\natexlab#1{#1}\fi

\bibitem[{{Bagetakos} {et~al.}(2011){Bagetakos}, {Brinks}, {Walter}, {de Blok},
  {Usero}, {Leroy}, {Rich}, \& {Kennicutt}}]{2011AJ....141...23B}
{Bagetakos}, I., {Brinks}, E., {Walter}, F., {et~al.} 2011, \aj, 141, 23

\bibitem[{{Bajaja} {et~al.}(2005){Bajaja}, {Arnal}, {Larrarte}, {Morras},
  {P{\"o}ppel}, \& {Kalberla}}]{2005A&A...440..767B}
{Bajaja}, E., {Arnal}, E.~M., {Larrarte}, J.~J., {et~al.} 2005, \aap, 440, 767

\bibitem[{{Brand} \& {Blitz}(1993)}]{1993A&A...275...67B}
{Brand}, J. \& {Blitz}, L. 1993, \aap, 275, 67

\bibitem[{{Daigle} {et~al.}(2007){Daigle}, {Joncas}, \&
  {Parizeau}}]{2007ApJ...661..285D}
{Daigle}, A., {Joncas}, G., \& {Parizeau}, M. 2007, \apj, 661, 285

\bibitem[{{Deharveng} {et~al.}(2010){Deharveng}, {Schuller}, {Anderson},
  {Zavagno}, {Wyrowski}, {Menten}, {Bronfman}, {Testi}, {Walmsley}, \&
  {Wienen}}]{2010A&A...523A...6D}
{Deharveng}, L., {Schuller}, F., {Anderson}, L.~D., {et~al.} 2010, \aap, 523,
  A6

\bibitem[{{Ehlerov{\'a}} \& {Palou{\v s}}(2005)}]{2005A&A...437..101E}
{Ehlerov{\'a}}, S. \& {Palou{\v s}}, J. 2005, \aap, 437, 101, paper 1

\bibitem[{{Elmegreen}(2012)}]{2012IAUS..284..317E}
{Elmegreen}, B.~G. 2012, in IAU Symposium, Vol. 284, IAU Symposium, 317--329

\bibitem[{{Gooch}(1997)}]{1997PASA...14..106G}
{Gooch}, R.~E. 1997, \pasa, 14, 106

\bibitem[{{Hartmann} \& {Burton}(1997)}]{1997agnh.book.....H}
{Hartmann}, D. \& {Burton}, W.~B. 1997, {Atlas of Galactic Neutral Hydrogen}

\bibitem[{{Heiles}(1979)}]{1979ApJ...229..533H}
{Heiles}, C. 1979, \apj, 229, 533

\bibitem[{{Kalberla} {et~al.}(2005){Kalberla}, {Burton}, {Hartmann}, {Arnal},
  {Bajaja}, {Morras}, \& {P{\"o}ppel}}]{2005A&A...440..775K}
{Kalberla}, P.~M.~W., {Burton}, W.~B., {Hartmann}, D., {et~al.} 2005, \aap,
  440, 775

\bibitem[{{Kalberla} \& {Dedes}(2008)}]{2008A&A...487..951K}
{Kalberla}, P.~M.~W. \& {Dedes}, L. 2008, \aap, 487, 951

\bibitem[{{Kalberla} {et~al.}(2010){Kalberla}, {McClure-Griffiths}, {Pisano},
  {Calabretta}, {Ford}, {Lockman}, {Staveley-Smith}, {Kerp}, {Winkel},
  {Murphy}, \& {Newton-McGee}}]{2010A&A...521A..17K}
{Kalberla}, P.~M.~W., {McClure-Griffiths}, N.~M., {Pisano}, D.~J., {et~al.}
  2010, \aap, 521, A17

\bibitem[{{Kerp} {et~al.}(2011){Kerp}, {Winkel}, {Ben Bekhti}, {Fl{\"o}er}, \&
  {Kalberla}}]{2011AN....332..637K}
{Kerp}, J., {Winkel}, B., {Ben Bekhti}, N., {Fl{\"o}er}, L., \& {Kalberla},
  P.~M.~W. 2011, Astronomische Nachrichten, 332, 637

\bibitem[{{Levine} {et~al.}(2006){Levine}, {Blitz}, \&
  {Heiles}}]{2006ApJ...643..881L}
{Levine}, E.~S., {Blitz}, L., \& {Heiles}, C. 2006, \apj, 643, 881

\bibitem[{{McClure-Griffiths} {et~al.}(2002){McClure-Griffiths}, {Dickey},
  {Gaensler}, \& {Green}}]{2002ApJ...578..176M}
{McClure-Griffiths}, N.~M., {Dickey}, J.~M., {Gaensler}, B.~M., \& {Green},
  A.~J. 2002, \apj, 578, 176

\bibitem[{{McKee} \& {Williams}(1997)}]{1997ApJ...476..144M}
{McKee}, C.~F. \& {Williams}, J.~P. 1997, \apj, 476, 144

\bibitem[{{Oey} \& {Clarke}(1997)}]{1997MNRAS.289..570O}
{Oey}, M.~S. \& {Clarke}, C.~J. 1997, \mnras, 289, 570

\bibitem[{{Thilker} {et~al.}(1998){Thilker}, {Braun}, \&
  {Walterbos}}]{1998A&A...332..429T}
{Thilker}, D.~A., {Braun}, R., \& {Walterbos}, R.~M. 1998, \aap, 332, 429

\bibitem[{{Weaver} {et~al.}(1977){Weaver}, {McCray}, {Castor}, {Shapiro}, \&
  {Moore}}]{1977ApJ...218..377W}
{Weaver}, R., {McCray}, R., {Castor}, J., {Shapiro}, P., \& {Moore}, R. 1977,
  \apj, 218, 377

\bibitem[{{Zavagno} {et~al.}(2006){Zavagno}, {Deharveng}, {Comer{\'o}n},
  {Brand}, {Massi}, {Caplan}, \& {Russeil}}]{2006A&A...446..171Z}
{Zavagno}, A., {Deharveng}, L., {Comer{\'o}n}, F., {et~al.} 2006, \aap, 446,
  171

\end{thebibliography}

\Online

\begin{appendix}

\section{Online tables}

Table \ref{table1_onl} contains observed properties of HI shells. Column 1 is
a running number of the structure, Col. 2 is the name of structure (GSH +
galactic coordinates of the centre and the radial velocity). 
The remaining columns describe the ranges in coordinates,
where the structure is visible:  Col. 3 and Col. 4 the longitude,
Col. 5 and Col. 6 the latitude, Col. 7 and Col. 8 the radial velocity.

Table \ref{table2_onl} contains derived properties of HI shells.
Column 1 is a running number of the structure, Col. 2 is the name of the
structure (GSH + galactic coordinates of the centre and the radial velocity).
Col. 3 is the galactocentric distance of the structure 
\citep[using the rotation curve of][]{1993A&A...275...67B}, 
Col. 4 is the radius of the shell 
(calculated as the geometric mean of dimensions in $l$ and $b$), Col. 5 is 
the expansion
velocity (calculated as a half of the velocity extent of the structure).
Columns 6 and 7 are estimates of the needed energy input into the shell
and the evolutionary time based on the model of \citet{1977ApJ...218..377W}.
Shells in the inner Galaxy are excluded from this table, as are
shells found at very low velocities 
$|v_{\mathrm{lsr}}| < 10\ \mathrm{kms}^{-1}$, 
shells close to the centre and anticentre directions
($l < 20^{\degr}$, $l > 340^{\degr}$, $170^{\degr} < l < 190^{\degr}$),
and shells with very high $b$ coordinates ($|b| > 50^{\degr}$).

{\onecolumn
\begin{longtable}{|rlrrrrrr|}
 \caption{\label{table1_onl}Observed properties of HI shells
 (designation of the structure in the form of GSH, galactic $l$ and $b$
 coordinates, the radial velocity $v$; ranges in coordinates and
 velocities where the structure is visible).}\\
 \hline
 Num & name & 
 $l_1$ & $l_2$ & $b_1$ & $b_2$ & $v_1$ & $v_2$  \\
 \hline
 \endfirsthead
 \caption{continued.}\\
 \hline
 Num & name & 
 $l_1$ & $l_2$ & $b_1$ & $b_2$ & $v_1$ & $v_2$  \\
 \hline
 \endhead
 \hline
 \endfoot
 \hline
 \endlastfoot
      1 & GSH002.0+23.5+013.4  &     0.5 &     3.5 &    22.5 &    24.5 &     8.2 &    18.6 \\
     2 & GSH005.0+14.0-005.2  &   348.5 &    25.0 &     0.0 &    31.0 &   -25.8 &    15.5 \\
     3 & GSH005.0-06.5+010.3  &     1.0 &     8.0 &    -9.5 &    -4.0 &     7.2 &    14.4 \\
     4 & GSH005.5-27.5+015.5  &     1.0 &     8.5 &   -29.5 &   -25.5 &     9.3 &    21.6 \\
     5 & GSH007.5+19.0+010.3  &     4.0 &    11.5 &    17.5 &    22.0 &     3.1 &    19.6 \\
     6 & GSH007.5-02.0+057.7  &     6.5 &     9.0 &    -3.0 &    -1.5 &    49.5 &    63.9 \\
     7 & GSH008.0-05.5-001.0  &     1.0 &    12.5 &    -8.5 &    -2.5 &    -8.2 &     7.2 \\
     8 & GSH010.5+03.5+044.3  &     9.5 &    12.0 &     3.0 &     4.5 &    40.2 &    49.5 \\
     9 & GSH010.5+08.0+013.4  &     8.5 &    12.5 &     5.0 &    10.0 &     6.2 &    21.6 \\
    10 & GSH012.0+13.0+012.4  &     7.5 &    17.0 &     8.5 &    15.5 &     6.2 &    22.7 \\
    11 & GSH013.0+02.5+011.3  &    10.5 &    15.5 &     0.5 &     5.0 &     6.2 &    17.5 \\
    12 & GSH014.5+28.0+017.5  &   354.0 &    32.5 &    18.0 &    37.5 &     1.0 &    30.9 \\
    13 & GSH019.5+13.5-007.2  &    17.5 &    21.0 &    12.0 &    15.0 &   -12.4 &    -3.1 \\
    14 & GSH021.5+06.5+027.8  &    20.5 &    22.0 &     5.5 &     7.5 &    22.7 &    33.0 \\
    15 & GSH022.5+03.5-022.7  &    21.5 &    23.5 &     2.5 &     4.5 &   -30.9 &   -13.4 \\
    16 & GSH026.0+01.5-044.3  &    25.5 &    28.5 &     1.0 &     3.0 &   -48.4 &   -41.2 \\
    17 & GSH026.0+03.5+015.5  &    21.0 &    32.5 &    -1.5 &     8.5 &     5.2 &    26.8 \\
    18 & GSH027.0-06.0+006.2  &    20.0 &    35.5 &   -13.5 &    -1.0 &    -7.2 &    17.5 \\
    19 & GSH027.5+03.5-026.8  &    26.0 &    29.5 &     2.5 &     5.5 &   -30.9 &   -22.7 \\
    20 & GSH028.5+02.0+042.3  &    27.0 &    30.0 &     1.0 &     3.0 &    38.1 &    48.4 \\
    21 & GSH029.0-04.5+027.8  &    26.5 &    31.5 &    -6.0 &    -3.0 &    21.6 &    31.9 \\
    22 & GSH030.5+15.5+015.5  &    25.5 &    34.0 &    11.0 &    20.0 &     7.2 &    23.7 \\
    23 & GSH034.5-02.0+004.1  &    31.5 &    37.0 &    -5.0 &     2.5 &    -4.1 &    13.4 \\
    24 & GSH035.0-06.5+052.6  &    32.5 &    36.5 &    -8.5 &    -4.0 &    34.0 &    64.9 \\
    25 & GSH037.5+09.5+008.2  &    32.5 &    41.5 &     6.0 &    12.5 &    -0.0 &    13.4 \\
    26 & GSH038.5+32.0+009.3  &    35.5 &    45.0 &    27.0 &    34.5 &     4.1 &    15.5 \\
    27 & GSH041.5-34.0+016.5  &    38.0 &    44.0 &   -37.0 &   -32.5 &     9.3 &    22.7 \\
    28 & GSH043.5+04.5+025.8  &    42.0 &    44.5 &     2.5 &     5.0 &    22.7 &    29.9 \\
    29 & GSH046.0+10.5+037.1  &    44.0 &    48.0 &     8.0 &    12.5 &    27.8 &    43.3 \\
    30 & GSH046.0-10.0+027.8  &    43.5 &    48.0 &   -13.0 &    -9.0 &    21.6 &    33.0 \\
    31 & GSH046.5+19.5-004.1  &    44.5 &    49.0 &    17.5 &    21.5 &   -12.4 &     1.0 \\
    32 & GSH046.5+18.0+006.2  &    43.5 &    49.0 &    16.5 &    20.0 &     3.1 &    10.3 \\
    33 & GSH046.5-02.5+039.2  &    45.5 &    47.0 &    -3.5 &    -1.5 &    34.0 &    44.3 \\
    34 & GSH047.5-09.5+014.4  &    46.5 &    49.5 &   -11.0 &    -8.0 &     9.3 &    19.6 \\
    35 & GSH047.5-32.5+004.1  &    45.0 &    49.5 &   -35.5 &   -29.0 &    -0.0 &     8.2 \\
    36 & GSH048.0-22.0-001.0  &   250.5 &   179.5 &   -53.0 &    16.0 &   -44.3 &    41.2 \\
    37 & GSH048.5-01.5-014.4  &    46.0 &    51.5 &    -2.5 &     0.0 &   -18.6 &    -9.3 \\
    38 & GSH049.0+09.0+013.4  &    41.5 &    56.5 &     5.5 &    12.5 &     3.1 &    30.9 \\
    39 & GSH049.0-00.5-027.8  &    47.0 &    50.5 &    -1.5 &     0.5 &   -33.0 &   -22.7 \\
    40 & GSH050.0-05.0+019.6  &    41.5 &    56.5 &    -8.5 &    -1.0 &    10.3 &    36.1 \\
    41 & GSH056.5+00.0-022.7  &    54.5 &    58.5 &    -0.5 &     1.0 &   -27.8 &   -15.5 \\
    42 & GSH058.0+09.0+029.9  &    57.0 &    60.5 &     7.5 &    11.0 &    23.7 &    35.0 \\
    43 & GSH059.0+09.0+015.5  &    58.5 &    59.5 &     8.5 &    10.0 &    11.3 &    19.6 \\
    44 & GSH061.5-00.5+046.4  &    61.0 &    62.5 &    -1.5 &     1.0 &    37.1 &    54.6 \\
    45 & GSH063.5-00.5-001.0  &    62.0 &    64.5 &    -2.0 &     0.5 &    -7.2 &     5.2 \\
    46 & GSH063.5+16.0+013.4  &    59.5 &    66.5 &    13.0 &    18.0 &    10.3 &    19.6 \\
    47 & GSH063.5+03.5+016.5  &    62.0 &    65.0 &     2.5 &     4.5 &    11.3 &    19.6 \\
    48 & GSH064.0-24.5+011.3  &    62.0 &    66.0 &   -26.5 &   -23.0 &     7.2 &    14.4 \\
    49 & GSH064.5-03.5-053.6  &    63.5 &    66.0 &    -4.5 &    -1.5 &   -57.7 &   -50.5 \\
    50 & GSH065.5-02.5+010.3  &    63.5 &    67.0 &    -3.5 &    -1.0 &     7.2 &    14.4 \\
    51 & GSH066.0-01.5-088.6  &    62.0 &    72.5 &    -3.5 &     0.0 &   -97.9 &   -77.3 \\
    52 & GSH068.0+02.0-079.4  &    67.0 &    69.0 &     1.0 &     3.0 &   -83.5 &   -76.3 \\
    53 & GSH068.5-03.0-043.3  &    66.5 &    75.0 &    -5.0 &    -0.5 &   -54.6 &   -29.9 \\
    54 & GSH069.5+09.0+014.4  &    67.5 &    70.5 &     6.5 &    11.0 &     9.3 &    20.6 \\
    55 & GSH070.0-00.5-027.8  &    69.0 &    71.0 &    -1.5 &     0.5 &   -34.0 &   -20.6 \\
    56 & GSH072.5-07.0+002.1  &    71.0 &    75.0 &    -8.0 &    -5.0 &    -4.1 &     7.2 \\
    57 & GSH073.0-09.0+016.5  &    71.5 &    74.5 &   -11.5 &    -7.5 &     9.3 &    21.6 \\
    58 & GSH074.0+04.0+028.9  &    73.0 &    75.5 &     3.0 &     5.5 &    24.7 &    31.9 \\
    59 & GSH074.0+06.0-002.1  &    71.5 &    76.0 &     4.5 &     8.0 &    -8.2 &     2.1 \\
    60 & GSH075.5-01.0+025.8  &    70.5 &    81.0 &    -4.0 &     2.5 &    11.3 &    37.1 \\
    61 & GSH077.0-05.5+014.4  &    75.5 &    78.5 &    -6.5 &    -4.5 &     9.3 &    21.6 \\
    62 & GSH077.0+02.5-039.2  &    76.0 &    78.0 &     1.5 &     3.0 &   -45.3 &   -34.0 \\
    63 & GSH077.0-07.0+002.1  &    76.0 &    78.0 &    -8.5 &    -6.0 &    -2.1 &     6.2 \\
    64 & GSH077.5+07.0-016.5  &    76.5 &    78.0 &     6.0 &     8.0 &   -19.6 &   -12.4 \\
    65 & GSH081.0-09.5+003.1  &    73.0 &    86.5 &   -14.5 &    -6.0 &   -19.6 &    25.8 \\
    66 & GSH083.0+03.5-053.6  &    81.5 &    84.0 &     3.0 &     4.5 &   -56.7 &   -49.5 \\
    67 & GSH083.5+01.0-023.7  &    81.0 &    87.5 &    -0.5 &     3.0 &   -29.9 &   -13.4 \\
    68 & GSH084.0+10.5+017.5  &    80.5 &    87.0 &     6.5 &    13.0 &     2.1 &    29.9 \\
    69 & GSH085.5-02.0-024.7  &    84.0 &    87.5 &    -3.0 &    -0.5 &   -28.9 &   -19.6 \\
    70 & GSH088.0-08.5-019.6  &    86.5 &    89.0 &    -9.5 &    -7.0 &   -24.7 &   -15.5 \\
    71 & GSH088.0+01.0-084.5  &    87.0 &    89.0 &    -0.5 &     2.5 &   -88.6 &   -80.4 \\
    72 & GSH088.5+01.0-097.9  &    86.5 &    90.0 &    -1.5 &     2.5 &  -103.1 &   -91.7 \\
    73 & GSH088.5-03.0+020.6  &    86.5 &    90.5 &    -5.0 &    -0.5 &    11.3 &    26.8 \\
    74 & GSH089.0-02.0-002.1  &    83.5 &    95.5 &    -9.0 &     5.5 &   -25.8 &    27.8 \\
    75 & GSH089.0+01.5-115.4  &    87.0 &    90.5 &    -0.5 &     3.5 &  -123.7 &  -107.2 \\
    76 & GSH091.0-05.0-059.8  &    88.0 &    94.0 &    -8.0 &    -2.0 &   -84.5 &   -25.8 \\
    77 & GSH091.0-08.5-001.0  &    89.0 &    93.0 &   -10.0 &    -7.5 &    -4.1 &     3.1 \\
    78 & GSH092.0+02.5-023.7  &    89.0 &    95.5 &    -0.5 &     6.0 &   -29.9 &   -17.5 \\
    79 & GSH092.0-05.5+004.1  &    88.5 &    94.5 &    -7.5 &    -4.0 &    -2.1 &     9.3 \\
    80 & GSH092.5-14.5-024.7  &    86.5 &    99.0 &   -18.5 &   -10.0 &   -40.2 &    -5.2 \\
    81 & GSH093.5-20.0+011.3  &    85.5 &   104.0 &   -28.5 &   -14.0 &     4.1 &    17.5 \\
    82 & GSH093.5+18.0-017.5  &    89.5 &    97.5 &    12.5 &    23.5 &   -34.0 &     1.0 \\
    83 & GSH094.0+10.5-011.3  &    92.5 &    95.5 &     9.5 &    13.0 &   -18.6 &    -5.2 \\
    84 & GSH096.0+11.5+010.3  &    94.0 &    97.5 &    10.0 &    14.0 &     6.2 &    14.4 \\
    85 & GSH096.5-03.5-028.9  &    95.0 &   100.0 &    -4.0 &    -2.5 &   -33.0 &   -25.8 \\
    86 & GSH096.5+03.5-113.4  &    90.5 &   105.0 &     0.5 &     5.5 &  -123.7 &  -103.1 \\
    87 & GSH097.0-11.5-011.3  &    96.0 &    99.0 &   -13.0 &    -9.0 &   -17.5 &    -5.2 \\
    88 & GSH098.0-16.5+004.1  &    89.5 &   104.0 &   -20.0 &   -12.5 &    -3.1 &    12.4 \\
    89 & GSH098.0-00.5+008.2  &    95.5 &    99.5 &    -4.0 &     2.5 &     2.1 &    13.4 \\
    90 & GSH099.5-11.0-022.7  &    97.5 &   102.0 &   -13.5 &    -8.0 &   -27.8 &   -13.4 \\
    91 & GSH100.0-23.5-007.2  &    96.5 &   105.0 &   -27.0 &   -19.5 &   -10.3 &    -2.1 \\
    92 & GSH100.5-02.0-038.1  &    98.0 &   102.0 &    -3.5 &    -1.0 &   -44.3 &   -33.0 \\
    93 & GSH100.5+03.0-029.9  &    96.5 &   106.5 &     0.5 &     5.5 &   -41.2 &   -20.6 \\
    94 & GSH102.0+08.0-017.5  &    91.0 &   112.5 &     3.0 &    12.0 &   -34.0 &    -0.0 \\
    95 & GSH103.5+04.0+001.0  &   100.5 &   106.0 &     2.0 &     7.5 &    -9.3 &    11.3 \\
    96 & GSH104.5-02.0-046.4  &   103.0 &   106.0 &    -3.0 &    -1.0 &   -52.6 &   -41.2 \\
    97 & GSH105.0-08.0-013.4  &    96.0 &   114.0 &   -14.5 &    -2.5 &   -41.2 &     4.1 \\
    98 & GSH105.0-04.0-057.7  &   104.0 &   106.0 &    -5.5 &    -2.0 &   -61.8 &   -53.6 \\
    99 & GSH106.0-07.0+009.3  &   102.0 &   110.5 &   -10.0 &    -4.5 &     3.1 &    18.6 \\
   100 & GSH106.5+13.0-043.3  &   105.5 &   107.5 &    12.0 &    14.0 &   -48.4 &   -38.1 \\
   101 & GSH107.0+00.5-079.4  &   104.5 &   108.0 &    -0.5 &     1.5 &   -82.4 &   -75.2 \\
   102 & GSH107.5+01.5-034.0  &   102.5 &   111.0 &    -1.0 &     4.0 &   -41.2 &   -27.8 \\
   103 & GSH107.5+22.5-006.2  &   104.0 &   110.0 &    21.0 &    25.0 &    -9.3 &    -2.1 \\
   104 & GSH108.0-01.5+007.2  &   104.5 &   112.0 &    -3.5 &     0.5 &    -0.0 &    15.5 \\
   105 & GSH108.0+06.0+014.4  &   106.0 &   111.5 &     4.5 &     7.5 &     9.3 &    19.6 \\
   106 & GSH108.5+06.5-035.0  &   106.5 &   111.0 &     4.5 &     7.5 &   -42.3 &   -28.9 \\
   107 & GSH109.0-08.0-060.8  &   107.5 &   111.0 &    -9.0 &    -6.5 &   -67.0 &   -54.6 \\
   108 & GSH109.5-04.5-063.9  &   108.0 &   111.0 &    -5.5 &    -3.0 &   -68.0 &   -58.7 \\
   109 & GSH109.5+17.5-018.6  &   106.5 &   113.5 &    15.0 &    19.5 &   -28.9 &    -9.3 \\
   110 & GSH110.5+08.5-047.4  &   109.0 &   112.5 &     7.0 &    10.0 &   -55.7 &   -41.2 \\
   111 & GSH110.5+03.5-006.2  &   107.5 &   113.5 &     1.0 &     7.5 &   -14.4 &    -0.0 \\
   112 & GSH111.5+01.0-079.4  &   109.5 &   114.5 &     0.0 &     2.0 &   -84.5 &   -75.2 \\
   113 & GSH111.5+01.5-101.0  &   110.5 &   112.5 &     0.5 &     2.0 &  -106.1 &   -96.9 \\
   114 & GSH117.0-04.0-026.8  &   114.0 &   119.5 &    -7.0 &    -1.0 &   -31.9 &   -19.6 \\
   115 & GSH118.5+06.5-104.1  &   115.0 &   123.0 &     4.5 &     8.5 &  -109.2 &   -96.9 \\
   116 & GSH119.5+04.0-095.8  &   115.0 &   122.0 &     2.5 &     5.0 &  -101.0 &   -86.6 \\
   117 & GSH121.5-05.0-037.1  &   115.5 &   124.5 &    -9.5 &     2.5 &   -61.8 &   -22.7 \\
   118 & GSH121.5+14.5-057.7  &   120.5 &   123.0 &    13.5 &    16.0 &   -63.9 &   -52.6 \\
   119 & GSH123.5-10.5-055.7  &   122.0 &   125.0 &   -12.5 &    -8.5 &   -60.8 &   -47.4 \\
   120 & GSH123.5+06.5-021.6  &   110.0 &   143.0 &    -3.5 &    25.5 &   -57.7 &     1.0 \\
   121 & GSH124.0+08.0+003.1  &   117.0 &   136.0 &     3.0 &    14.0 &    -4.1 &     9.3 \\
   122 & GSH125.5-08.5-014.4  &   123.5 &   128.0 &   -10.5 &    -6.0 &   -19.6 &   -11.3 \\
   123 & GSH127.5+16.5+008.2  &   125.5 &   131.5 &    13.0 &    19.0 &     4.1 &    12.4 \\
   124 & GSH127.5+00.5-105.1  &   124.0 &   134.5 &    -1.5 &     2.5 &  -112.3 &   -95.8 \\
   125 & GSH127.5+18.0-001.0  &   125.5 &   130.5 &    16.0 &    20.0 &    -5.2 &     4.1 \\
   126 & GSH128.5-15.5-026.8  &   125.5 &   133.0 &   -19.5 &   -12.0 &   -37.1 &   -13.4 \\
   127 & GSH128.5+03.5-033.0  &   127.0 &   129.5 &     2.5 &     5.0 &   -37.1 &   -29.9 \\
   128 & GSH129.5+15.0-017.5  &   127.0 &   132.0 &    12.0 &    17.0 &   -23.7 &   -10.3 \\
   129 & GSH129.5+06.5-067.0  &   125.5 &   132.0 &     5.0 &     7.5 &   -78.3 &   -57.7 \\
   130 & GSH130.5+04.5-051.5  &   128.0 &   132.5 &     3.5 &     6.5 &   -58.7 &   -41.2 \\
   131 & GSH131.5-13.0-021.6  &   130.5 &   133.5 &   -15.0 &   -11.5 &   -26.8 &   -16.5 \\
   132 & GSH132.5+15.5+003.1  &   130.5 &   134.5 &    12.5 &    17.5 &    -1.0 &     7.2 \\
   133 & GSH134.0-48.0-003.1  &   128.0 &   138.0 &   -50.5 &   -45.0 &    -7.2 &    -0.0 \\
   134 & GSH135.5-08.5-034.0  &   124.5 &   147.5 &   -13.5 &    -3.0 &   -63.9 &   -11.3 \\
   135 & GSH136.0+07.0-030.9  &   133.0 &   139.0 &     5.5 &     9.5 &   -38.1 &   -20.6 \\
   136 & GSH136.0+05.5-073.2  &   134.0 &   137.5 &     4.5 &     6.0 &   -77.3 &   -69.0 \\
   137 & GSH137.5-09.0-003.1  &   133.5 &   142.0 &   -11.5 &    -7.0 &    -7.2 &     2.1 \\
   138 & GSH138.0+16.0-008.2  &   134.0 &   141.5 &    12.5 &    21.5 &   -18.6 &    -2.1 \\
   139 & GSH138.5+10.0+011.3  &   131.0 &   148.0 &     6.5 &    14.0 &     3.1 &    18.6 \\
   140 & GSH138.5+19.5+007.2  &   131.5 &   147.0 &    15.5 &    22.5 &    -0.0 &    11.3 \\
   141 & GSH139.0-03.0-069.0  &   128.5 &   144.5 &    -5.5 &     0.0 &   -81.4 &   -59.8 \\
   142 & GSH139.0+35.5+002.1  &   124.5 &   163.0 &    22.5 &    51.0 &   -18.6 &    16.5 \\
   143 & GSH143.0+10.5-031.9  &   138.5 &   148.5 &     7.5 &    13.5 &   -46.4 &   -15.5 \\
   144 & GSH144.0+02.5-050.5  &   136.0 &   148.5 &    -0.5 &     5.5 &   -63.9 &   -37.1 \\
   145 & GSH145.0-03.5-010.3  &   140.5 &   149.5 &    -6.5 &    -0.5 &   -14.4 &    -5.2 \\
   146 & GSH145.0-52.5-015.5  &   141.5 &   150.0 &   -54.0 &   -51.0 &   -19.6 &   -12.4 \\
   147 & GSH145.0-10.0-063.9  &   143.5 &   147.0 &   -11.0 &    -9.0 &   -68.0 &   -59.8 \\
   148 & GSH145.5+40.0+009.3  &   144.0 &   148.5 &    38.5 &    41.5 &     5.2 &    12.4 \\
   149 & GSH147.5-09.0-038.1  &   145.0 &   149.0 &   -10.0 &    -8.0 &   -43.3 &   -35.0 \\
   150 & GSH147.5+16.0-004.1  &   142.0 &   155.0 &    12.5 &    20.5 &    -9.3 &    -0.0 \\
   151 & GSH148.0+20.5+007.2  &   145.5 &   151.0 &    18.0 &    22.5 &     2.1 &     9.3 \\
   152 & GSH150.0-12.0-017.5  &   148.5 &   151.5 &   -13.0 &   -11.5 &   -21.6 &   -14.4 \\
   153 & GSH151.0-04.0-049.5  &   145.5 &   155.0 &    -7.5 &     0.0 &   -61.8 &   -30.9 \\
   154 & GSH154.5-32.5+006.2  &   151.0 &   157.5 &   -34.0 &   -30.5 &     2.1 &     9.3 \\
   155 & GSH155.0+13.0-016.5  &   147.5 &   161.0 &    11.0 &    16.0 &   -26.8 &    -6.2 \\
   156 & GSH156.0-05.5-058.7  &   154.0 &   159.5 &    -7.0 &    -5.0 &   -64.9 &   -54.6 \\
   157 & GSH159.0-00.5-053.6  &   156.0 &   162.0 &    -2.5 &     0.5 &   -57.7 &   -50.5 \\
   158 & GSH160.0-38.0-002.1  &   152.0 &   165.0 &   -43.0 &   -35.0 &    -9.3 &     6.2 \\
   159 & GSH160.5+06.5-042.3  &   157.5 &   162.5 &     3.5 &     9.0 &   -47.4 &   -36.1 \\
   160 & GSH164.0+00.5-016.5  &   159.0 &   168.0 &    -1.0 &     2.5 &   -22.7 &   -11.3 \\
   161 & GSH167.0-10.5-017.5  &   160.0 &   171.5 &   -13.5 &    -9.0 &   -20.6 &   -13.4 \\
   162 & GSH167.5+14.0-003.1  &   164.0 &   172.0 &    11.5 &    17.0 &    -8.2 &    -0.0 \\
   163 & GSH168.0+11.5-023.7  &   167.0 &   170.5 &     9.5 &    12.5 &   -31.9 &   -16.5 \\
   164 & GSH168.5+22.5+015.5  &   163.5 &   172.0 &    19.5 &    27.5 &    11.3 &    19.6 \\
   165 & GSH173.0-07.0-028.9  &   170.5 &   175.5 &    -8.0 &    -5.5 &   -33.0 &   -23.7 \\
   166 & GSH173.5-05.0+010.3  &   167.0 &   175.5 &    -8.0 &    -2.5 &     5.2 &    17.5 \\
   167 & GSH175.0+00.5+014.4  &   174.0 &   176.5 &    -0.5 &     1.0 &     8.2 &    19.6 \\
   168 & GSH175.0+19.5-017.5  &   167.0 &   184.5 &    16.0 &    24.0 &   -30.9 &    -7.2 \\
   169 & GSH175.0-01.0-008.2  &   173.0 &   178.0 &    -2.0 &    -0.5 &   -14.4 &    -4.1 \\
   170 & GSH175.5-29.0+016.5  &   168.5 &   178.5 &   -33.5 &   -25.0 &     8.2 &    23.7 \\
   171 & GSH176.0+08.0-020.6  &   171.5 &   180.0 &     7.0 &     9.0 &   -24.7 &   -17.5 \\
   172 & GSH176.0+02.5+001.0  &   174.5 &   177.5 &     1.5 &     3.5 &    -3.1 &     4.1 \\
   173 & GSH179.0-47.5-000.0  &   168.5 &   189.0 &   -52.0 &   -42.5 &   -10.3 &    10.3 \\
   174 & GSH179.5-06.0-011.3  &   176.5 &   182.0 &    -7.5 &    -4.5 &   -15.5 &    -7.2 \\
   175 & GSH179.5+25.5-000.0  &   168.0 &   189.0 &    18.0 &    34.5 &   -12.4 &    12.4 \\
   176 & GSH180.0-36.5+016.5  &   177.0 &   184.0 &   -38.0 &   -35.0 &    12.4 &    20.6 \\
   177 & GSH180.0-24.5-004.1  &   174.5 &   186.0 &   -29.0 &   -18.0 &   -20.6 &     8.2 \\
   178 & GSH183.0-11.0+003.1  &   181.0 &   185.0 &   -13.0 &    -7.0 &    -2.1 &     7.2 \\
   179 & GSH184.5-13.5+013.4  &   178.0 &   190.0 &   -18.0 &    -8.5 &     1.0 &    25.8 \\
   180 & GSH185.0-17.5+003.1  &   179.5 &   189.5 &   -20.0 &   -14.0 &    -1.0 &     6.2 \\
   181 & GSH188.0-37.5+016.5  &   186.0 &   189.5 &   -39.0 &   -35.0 &    13.4 &    20.6 \\
   182 & GSH189.0+15.0+007.2  &   179.0 &   197.0 &    10.5 &    21.5 &    -4.1 &    19.6 \\
   183 & GSH190.5+08.0-017.5  &   188.5 &   192.0 &     6.5 &     8.5 &   -21.6 &   -11.3 \\
   184 & GSH192.0+05.5-015.5  &   191.0 &   193.5 &     4.5 &     7.0 &   -21.6 &    -9.3 \\
   185 & GSH192.0+00.0-018.6  &   183.0 &   197.5 &    -5.0 &     4.5 &   -28.9 &     1.0 \\
   186 & GSH192.5-05.5+010.3  &   190.0 &   193.5 &    -7.0 &    -4.5 &     6.2 &    13.4 \\
   187 & GSH193.0-10.5-006.2  &   192.0 &   194.0 &   -12.0 &   -10.0 &   -11.3 &    -3.1 \\
   188 & GSH194.0-06.5-004.1  &   192.5 &   195.5 &    -8.0 &    -5.0 &    -9.3 &     1.0 \\
   189 & GSH196.0-18.0+007.2  &   194.5 &   197.0 &   -19.0 &   -15.5 &    -0.0 &    11.3 \\
   190 & GSH196.0-13.5+002.1  &   193.5 &   199.5 &   -15.5 &   -10.5 &    -3.1 &     6.2 \\
   191 & GSH197.5-10.0+003.1  &   192.0 &   203.0 &   -13.0 &    -5.5 &   -10.3 &    14.4 \\
   192 & GSH200.0-19.5+020.6  &   196.0 &   203.0 &   -23.0 &   -17.5 &    13.4 &    27.8 \\
   193 & GSH200.5+06.0+022.7  &   198.0 &   203.0 &     5.0 &     7.0 &    17.5 &    26.8 \\
   194 & GSH201.5-18.0+008.2  &   200.5 &   203.5 &   -19.0 &   -17.0 &     5.2 &    12.4 \\
   195 & GSH201.5+07.0-003.1  &   196.0 &   206.5 &    -1.0 &    14.0 &   -20.6 &    14.4 \\
   196 & GSH203.0-04.5+014.4  &   200.5 &   206.0 &    -7.5 &    -1.0 &     9.3 &    21.6 \\
   197 & GSH203.0-38.5+013.4  &   201.0 &   206.0 &   -39.5 &   -38.0 &     9.3 &    16.5 \\
   198 & GSH204.0-10.5+020.6  &   202.0 &   206.5 &   -12.5 &    -9.0 &    16.5 &    23.7 \\
   199 & GSH206.0-08.5+007.2  &   203.0 &   208.0 &   -10.5 &    -6.5 &     3.1 &    13.4 \\
   200 & GSH206.0+07.5-004.1  &   204.0 &   208.0 &     6.0 &     9.0 &   -11.3 &     1.0 \\
   201 & GSH206.0-19.5+021.6  &   203.5 &   207.5 &   -21.5 &   -18.5 &    17.5 &    24.7 \\
   202 & GSH208.0-03.0-000.0  &   207.0 &   209.0 &    -4.0 &    -1.5 &    -4.1 &     5.2 \\
   203 & GSH210.0-04.0-012.4  &   199.0 &   224.5 &   -20.5 &     6.5 &   -21.6 &     7.2 \\
   204 & GSH210.0+09.0+004.1  &   207.5 &   214.0 &     6.0 &    11.5 &    -2.1 &    11.3 \\
   205 & GSH210.0-20.0+020.6  &   205.5 &   214.0 &   -24.0 &   -14.5 &     7.2 &    34.0 \\
   206 & GSH210.5-04.5+029.9  &   208.0 &   213.0 &    -6.5 &    -2.0 &    23.7 &    35.0 \\
   207 & GSH210.5-15.0-004.1  &   209.5 &   212.5 &   -16.0 &   -14.0 &    -9.3 &     2.1 \\
   208 & GSH214.0+00.0+017.5  &   208.5 &   218.0 &    -2.5 &     2.5 &     8.2 &    23.7 \\
   209 & GSH215.5-03.0+053.6  &   214.0 &   217.0 &    -3.5 &    -2.0 &    48.4 &    59.8 \\
   210 & GSH217.0+02.0+061.8  &   215.5 &   218.5 &     1.0 &     3.0 &    57.7 &    66.0 \\
   211 & GSH217.0-06.0+026.8  &   213.0 &   219.5 &    -8.0 &    -2.0 &    14.4 &    43.3 \\
   212 & GSH218.5-09.0+029.9  &   217.0 &   221.0 &   -10.5 &    -7.5 &    23.7 &    35.0 \\
   213 & GSH218.5-39.0-000.0  &   166.5 &   275.5 &   -66.5 &   -15.5 &   -22.7 &    27.8 \\
   214 & GSH219.5-15.0+015.5  &   215.0 &   224.5 &   -19.0 &   -12.5 &     7.2 &    23.7 \\
   215 & GSH221.0-07.0+009.3  &   218.5 &   222.5 &    -8.0 &    -5.5 &     6.2 &    14.4 \\
   216 & GSH221.5+05.5+012.4  &   216.5 &   225.0 &     3.0 &     9.0 &     8.2 &    18.6 \\
   217 & GSH222.0-05.0+025.8  &   220.0 &   224.0 &    -7.0 &    -2.5 &    16.5 &    35.0 \\
   218 & GSH222.0-03.0+046.4  &   219.0 &   226.5 &    -5.0 &    -1.0 &    42.3 &    52.6 \\
   219 & GSH223.0+03.0+069.0  &   221.5 &   225.0 &     2.0 &     4.5 &    63.9 &    73.2 \\
   220 & GSH224.5-00.5+023.7  &   219.0 &   231.0 &    -4.5 &     4.0 &    14.4 &    38.1 \\
   221 & GSH226.5-09.5+031.9  &   221.0 &   231.5 &   -14.5 &    -4.5 &    11.3 &    54.6 \\
   222 & GSH227.0-13.5+012.4  &   226.0 &   228.0 &   -15.5 &   -12.5 &     7.2 &    17.5 \\
   223 & GSH228.5+02.5+066.0  &   226.5 &   230.5 &     1.5 &     3.5 &    60.8 &    71.1 \\
   224 & GSH230.5+54.5-013.4  &   228.5 &   232.5 &    52.0 &    56.0 &   -18.6 &    -7.2 \\
   225 & GSH235.0+05.5+028.9  &   234.0 &   236.0 &     4.5 &     6.5 &    25.8 &    33.0 \\
   226 & GSH238.5-02.0+068.0  &   237.5 &   240.0 &    -3.0 &    -1.0 &    64.9 &    72.1 \\
   227 & GSH243.5-02.5+043.3  &   232.0 &   251.5 &   -11.5 &     7.0 &    16.5 &    78.3 \\
   228 & GSH243.5-15.5+025.8  &   240.0 &   246.0 &   -18.0 &   -11.5 &    14.4 &    36.1 \\
   229 & GSH244.0-05.5+070.1  &   242.5 &   245.0 &    -7.0 &    -5.0 &    66.0 &    73.2 \\
   230 & GSH244.0+35.0+005.2  &   234.0 &   254.0 &    26.5 &    45.5 &    -6.2 &    14.4 \\
   231 & GSH244.0-09.0+035.0  &   240.0 &   249.0 &   -11.5 &    -7.5 &    27.8 &    42.3 \\
   232 & GSH247.0-24.0+015.5  &   244.5 &   249.5 &   -28.0 &   -19.5 &     7.2 &    20.6 \\
   233 & GSH250.0+01.5+067.0  &   248.5 &   252.0 &     0.5 &     2.5 &    62.9 &    71.1 \\
   234 & GSH251.0-07.0+048.4  &   250.0 &   252.5 &    -8.5 &    -6.0 &    43.3 &    53.6 \\
   235 & GSH251.0-23.0-000.0  &   248.5 &   254.0 &   -25.5 &   -20.0 &    -6.2 &     6.2 \\
   236 & GSH252.0-02.5+034.0  &   251.0 &   252.5 &    -3.5 &    -1.5 &    28.9 &    38.1 \\
   237 & GSH252.5+01.0+017.5  &   250.5 &   254.5 &    -1.5 &     3.0 &     8.2 &    27.8 \\
   238 & GSH253.0-08.0+013.4  &   251.0 &   254.5 &    -9.5 &    -6.5 &     9.3 &    17.5 \\
   239 & GSH254.0+05.5+021.6  &   251.5 &   257.5 &     4.0 &     7.0 &    13.4 &    29.9 \\
   240 & GSH254.5+00.0+056.7  &   252.0 &   257.0 &    -2.5 &     2.5 &    44.3 &    64.9 \\
   241 & GSH255.5+54.0-010.3  &   249.0 &   262.5 &    51.5 &    56.5 &   -15.5 &    -4.1 \\
   242 & GSH256.0-04.5+079.4  &   249.5 &   261.5 &    -6.5 &    -0.5 &    57.7 &   109.2 \\
   243 & GSH256.0-06.0-002.1  &   253.0 &   260.0 &   -11.0 &    -1.0 &   -10.3 &     9.3 \\
   244 & GSH256.5-07.5+035.0  &   255.0 &   258.5 &   -10.0 &    -4.5 &    23.7 &    49.5 \\
   245 & GSH258.0+09.5+038.1  &   249.5 &   266.0 &     5.5 &    15.0 &     8.2 &    68.0 \\
   246 & GSH259.0+03.0+016.5  &   257.0 &   260.5 &     1.5 &     4.5 &     9.3 &    21.6 \\
   247 & GSH259.5-01.5+021.6  &   258.0 &   261.0 &    -3.0 &    -0.5 &    18.6 &    25.8 \\
   248 & GSH260.5-04.5+064.9  &   259.0 &   262.0 &    -5.5 &    -4.0 &    60.8 &    68.0 \\
   249 & GSH261.0+02.5+030.9  &   260.0 &   263.0 &     1.0 &     4.0 &    24.7 &    38.1 \\
   250 & GSH262.5-01.5+041.2  &   259.0 &   265.5 &    -5.5 &     3.0 &    17.5 &    55.7 \\
   251 & GSH263.5-16.0-005.2  &   263.0 &   265.0 &   -17.5 &   -15.0 &    -8.2 &    -1.0 \\
   252 & GSH263.5+07.0+002.1  &   260.5 &   267.5 &     4.5 &    10.5 &    -1.0 &     8.2 \\
   253 & GSH263.5-11.5-004.1  &   262.0 &   265.0 &   -13.5 &   -10.0 &   -12.4 &     2.1 \\
   254 & GSH265.0-05.5+082.4  &   263.0 &   268.0 &    -7.0 &    -4.0 &    71.1 &    90.7 \\
   255 & GSH265.5-03.0-006.2  &   264.5 &   266.5 &    -5.0 &    -0.5 &   -11.3 &    -2.1 \\
   256 & GSH265.5-04.5+007.2  &   262.5 &   268.0 &    -6.5 &    -1.5 &     3.1 &    10.3 \\
   257 & GSH266.0+04.0+037.1  &   265.0 &   267.0 &     3.0 &     5.0 &    31.9 &    41.2 \\
   258 & GSH266.5-04.5+098.9  &   264.5 &   268.5 &    -8.0 &    -2.5 &    93.8 &   105.1 \\
   259 & GSH266.5+07.5+020.6  &   263.0 &   269.5 &     6.0 &     9.5 &     6.2 &    30.9 \\
   260 & GSH267.0-01.0+077.3  &   264.5 &   269.5 &    -3.0 &     1.0 &    69.0 &    85.5 \\
   261 & GSH267.5-03.0+114.4  &   265.0 &   270.5 &    -5.5 &    -1.5 &   106.1 &   121.6 \\
   262 & GSH268.0-04.0+030.9  &   267.0 &   269.0 &    -5.0 &    -3.0 &    26.8 &    34.0 \\
   263 & GSH268.5-11.0+001.0  &   266.5 &   271.5 &   -14.5 &    -8.0 &    -5.2 &     5.2 \\
   264 & GSH269.5-07.0+077.3  &   268.0 &   271.0 &    -8.0 &    -6.5 &    71.1 &    84.5 \\
   265 & GSH270.0-02.5+042.3  &   266.5 &   271.5 &    -5.0 &     1.0 &    23.7 &    68.0 \\
   266 & GSH270.0-05.5+096.9  &   268.5 &   271.5 &    -6.5 &    -4.5 &    92.8 &   100.0 \\
   267 & GSH270.5-09.0-006.2  &   267.5 &   274.0 &   -11.0 &    -7.5 &   -11.3 &    -1.0 \\
   268 & GSH271.0+11.5-014.4  &   269.0 &   272.0 &     9.5 &    13.0 &   -21.6 &    -7.2 \\
   269 & GSH271.5-02.0+103.1  &   270.0 &   273.5 &    -2.5 &    -1.0 &    98.9 &   107.2 \\
   270 & GSH272.0+04.0+024.7  &   269.0 &   274.5 &     2.5 &     6.5 &     8.2 &    44.3 \\
   271 & GSH273.0-26.5-004.1  &   271.0 &   275.0 &   -28.0 &   -25.0 &    -9.3 &     2.1 \\
   272 & GSH273.0+06.0+014.4  &   272.0 &   274.5 &     5.5 &     7.5 &    10.3 &    20.6 \\
   273 & GSH273.5-07.5+031.9  &   270.0 &   276.5 &   -11.5 &    -3.5 &     8.2 &    66.0 \\
   274 & GSH274.5-03.0-012.4  &   273.5 &   275.5 &    -4.0 &    -2.0 &   -18.6 &    -6.2 \\
   275 & GSH276.0+00.0+036.1  &   271.5 &   279.5 &    -3.0 &     2.5 &    13.4 &    53.6 \\
   276 & GSH276.5-09.5-001.0  &   272.0 &   281.5 &   -12.0 &    -5.5 &    -6.2 &     5.2 \\
   277 & GSH276.5+03.5-016.5  &   274.5 &   279.0 &     2.0 &     5.5 &   -26.8 &    -9.3 \\
   278 & GSH277.0-12.0-006.2  &   275.0 &   278.5 &   -13.0 &   -10.5 &   -11.3 &    -2.1 \\
   279 & GSH277.0+09.0+001.0  &   271.0 &   281.5 &     5.5 &    14.0 &   -11.3 &    18.6 \\
   280 & GSH278.5-40.0+008.2  &   276.0 &   281.5 &   -41.0 &   -39.0 &     4.1 &    11.3 \\
   281 & GSH278.5-02.5+026.8  &   277.0 &   279.5 &    -4.0 &    -0.5 &    18.6 &    33.0 \\
   282 & GSH280.5-05.0+071.1  &   277.0 &   283.5 &    -7.5 &    -3.5 &    60.8 &    78.3 \\
   283 & GSH281.5+06.0-015.5  &   279.5 &   283.5 &     3.5 &     7.5 &   -24.7 &    -7.2 \\
   284 & GSH281.5-01.5+085.5  &   277.0 &   284.0 &    -3.5 &    -0.5 &    79.4 &    91.7 \\
   285 & GSH283.5+37.5-008.2  &   277.0 &   290.5 &    31.0 &    42.0 &   -12.4 &    -3.1 \\
   286 & GSH283.5+34.0+001.0  &   280.5 &   285.0 &    31.5 &    37.0 &    -3.1 &     4.1 \\
   287 & GSH283.5+12.0+009.3  &   282.0 &   286.5 &    10.0 &    13.5 &    -2.1 &    19.6 \\
   288 & GSH284.5-05.5-002.1  &   282.0 &   286.5 &   -10.0 &    -2.5 &    -7.2 &     2.1 \\
   289 & GSH285.5-02.5+083.5  &   282.5 &   288.5 &    -5.0 &    -0.5 &    76.3 &    93.8 \\
   290 & GSH286.0+05.0-025.8  &   284.0 &   288.0 &     2.5 &     8.0 &   -37.1 &   -12.4 \\
   291 & GSH289.0-02.5-031.9  &   288.0 &   290.5 &    -4.0 &    -1.5 &   -40.2 &   -22.7 \\
   292 & GSH289.5+00.5-028.9  &   288.5 &   290.5 &    -1.0 &     2.5 &   -33.0 &   -24.7 \\
   293 & GSH289.5-36.0-007.2  &   285.5 &   296.0 &   -38.5 &   -32.5 &   -13.4 &    -0.0 \\
   294 & GSH292.0-01.5+050.5  &   290.5 &   294.0 &    -3.0 &     0.0 &    42.3 &    60.8 \\
   295 & GSH292.5-01.0+083.5  &   291.5 &   293.5 &    -1.5 &    -0.5 &    79.4 &    86.6 \\
   296 & GSH297.0-00.5+069.0  &   292.0 &   302.0 &    -3.0 &     1.5 &    55.7 &    85.5 \\
   297 & GSH300.0-03.0+082.4  &   298.5 &   301.5 &    -4.0 &    -2.0 &    79.4 &    86.6 \\
   298 & GSH301.5-03.0-054.6  &   300.5 &   303.0 &    -3.5 &    -2.5 &   -58.7 &   -48.4 \\
   299 & GSH303.5+07.5-036.1  &   302.5 &   304.5 &     6.0 &     9.0 &   -40.2 &   -31.9 \\
   300 & GSH304.0-01.0+061.8  &   302.5 &   305.5 &    -3.0 &     1.5 &    50.5 &    74.2 \\
   301 & GSH305.0-02.5-020.6  &   303.0 &   307.0 &    -4.0 &    -0.5 &   -28.9 &   -10.3 \\
   302 & GSH307.0-08.0-020.6  &   304.5 &   310.5 &   -11.5 &    -6.0 &   -23.7 &   -16.5 \\
   303 & GSH307.0-02.5-034.0  &   306.0 &   308.5 &    -3.5 &    -1.5 &   -37.1 &   -29.9 \\
   304 & GSH310.0-01.5+061.8  &   309.0 &   311.5 &    -3.0 &     0.5 &    55.7 &    70.1 \\
   305 & GSH316.0-03.5-014.4  &   313.5 &   318.5 &    -6.0 &    -2.0 &   -19.6 &    -9.3 \\
   306 & GSH316.5-29.0+001.0  &   313.5 &   318.5 &   -33.0 &   -27.5 &    -5.2 &     5.2 \\
   307 & GSH317.5+09.5-040.2  &   315.5 &   319.0 &     8.0 &    11.5 &   -55.7 &   -25.8 \\
   308 & GSH318.0-03.5-036.1  &   317.0 &   319.5 &    -5.5 &    -3.0 &   -41.2 &   -31.9 \\
   309 & GSH319.0-03.5+023.7  &   317.5 &   320.5 &    -5.0 &    -2.5 &    14.4 &    36.1 \\
   310 & GSH319.5+05.0+009.3  &   313.0 &   325.0 &     1.0 &     8.5 &     5.2 &    15.5 \\
   311 & GSH321.0-03.5-027.8  &   318.0 &   323.0 &    -5.0 &    -2.0 &   -35.0 &   -22.7 \\
   312 & GSH325.5+71.5-003.1  &   313.0 &   336.0 &    68.0 &    73.5 &    -6.2 &     1.0 \\
   313 & GSH326.5+07.0+001.0  &   318.0 &   334.5 &     2.5 &    10.5 &    -4.1 &     4.1 \\
   314 & GSH327.0+09.0+008.2  &   323.5 &   329.0 &     7.0 &    10.5 &     4.1 &    11.3 \\
   315 & GSH329.0+06.0-018.6  &   326.0 &   331.0 &     4.5 &     7.0 &   -21.6 &   -13.4 \\
   316 & GSH330.0-05.5-035.0  &   329.5 &   331.0 &    -6.5 &    -4.5 &   -39.2 &   -30.9 \\
   317 & GSH331.0-08.5+011.3  &   327.0 &   335.0 &   -10.5 &    -6.5 &    -0.0 &    22.7 \\
   318 & GSH332.0-09.5-007.2  &   330.0 &   335.5 &   -10.5 &    -7.5 &   -13.4 &    -2.1 \\
   319 & GSH333.0-03.5+001.0  &   323.0 &   340.5 &    -7.0 &    -0.5 &   -14.4 &    17.5 \\
   320 & GSH333.5+11.5+014.4  &   330.0 &   336.0 &     9.5 &    14.5 &     6.2 &    20.6 \\
   321 & GSH334.0+03.5-000.0  &   328.0 &   339.5 &    -0.5 &     8.0 &   -15.5 &    13.4 \\
   322 & GSH338.5-10.0+012.4  &   335.5 &   341.5 &   -12.0 &    -8.0 &     6.2 &    17.5 \\
   323 & GSH339.5+02.5+005.2  &   337.0 &   342.5 &     1.0 &     3.5 &     2.1 &     9.3 \\
   324 & GSH342.0-02.0-017.5  &   340.5 &   343.5 &    -2.5 &    -1.0 &   -23.7 &   -12.4 \\
   325 & GSH344.0-04.5-003.1  &   342.0 &   346.5 &    -7.5 &    -1.5 &    -7.2 &     1.0 \\
   326 & GSH344.5+04.5-002.1  &   342.5 &   346.5 &     1.5 &     7.0 &    -7.2 &     1.0 \\
   327 & GSH345.0-04.0+012.4  &   343.5 &   347.0 &    -5.0 &    -2.5 &     9.3 &    16.5 \\
   328 & GSH350.0+05.0+008.2  &   346.0 &   354.5 &     1.5 &     8.0 &   -14.4 &    27.8 \\
   329 & GSH350.0+11.0-000.0  &   346.0 &   353.5 &     8.5 &    14.0 &    -6.2 &     5.2 \\
   330 & GSH351.0-05.5+014.4  &   349.0 &   353.5 &    -7.5 &    -4.0 &     7.2 &    21.6 \\
   331 & GSH354.5+09.0+008.2  &   352.5 &   358.0 &     6.5 &    10.5 &     3.1 &    13.4 \\
   332 & GSH356.5+10.0-010.3  &   351.5 &     2.0 &     8.0 &    12.0 &   -19.6 &     1.0 \\
   333 & GSH359.5-28.5+006.2  &   356.5 &     4.0 &   -30.0 &   -27.0 &     1.0 &    11.3 \\

\end{longtable}
}

{\onecolumn
\begin{longtable}{|rlrrrrr|}
 \caption{\label{table2_onl}Derived properties of HI shells in the outer Galaxy
 (designation of the shell, galactocentric distance, radius of the
 shell, expansion velocity, estimated energy input, estimated evolutionary
 time).}\\
 \hline
 Num & name & $R\ (\mathrm{kpc})$ & $r_{\mathrm{sh}} (\mathrm{kpc})$ & 
 $v_{\mathrm{exp}} (\mathrm{kms}^{-1})$ &
 $L\ (10^{51} \mathrm{ergMyr}^{-1})$ & $t_{\mathrm{exp}} (\mathrm{Myr})$ \\
 \hline
 \endfirsthead
 \caption{continued.}\\
 \hline
 Num & name & $R\ (\mathrm{kpc})$ & $r_{\mathrm{sh}} (\mathrm{kpc})$ & 
 $v_{\mathrm{exp}} (\mathrm{kms}^{-1})$ &
 $L\ (10^{51} \mathrm{ergMyr}^{-1})$ & $t_{\mathrm{exp}} (\mathrm{Myr})$ \\
 \hline
 \endhead
 \hline
 \endfoot
      1 & GSH022.5+03.5-022.7  &  12.0 &  0.42 &   9.3 &   0.17E+00 &   26.8 \\
     2 & GSH026.0+01.5-044.3  &  16.3 &  0.61 &   4.1 &   0.52E-02 &   86.7 \\
     3 & GSH027.5+03.5-026.8  &  11.8 &  0.61 &   4.6 &   0.40E-01 &   77.2 \\
     4 & GSH048.5-01.5-014.4  &   9.5 &  0.47 &   5.2 &   0.35E-02 &   53.7 \\
     5 & GSH049.0-00.5-027.8  &  10.5 &  0.38 &   5.7 &   0.19E-01 &   39.7 \\
     6 & GSH056.5+00.0-022.7  &   9.9 &  0.30 &   6.7 &   0.95E-03 &   26.6 \\
     7 & GSH064.5-03.5-053.6  &  12.0 &  0.37 &   4.1 &   0.78E-02 &   52.0 \\
     8 & GSH066.0-01.5-088.6  &  15.8 &  1.00 &  10.8 &   0.11E+00 &   54.3 \\
     9 & GSH068.0+02.0-079.4  &  14.4 &  0.33 &   4.1 &   0.28E-02 &   47.5 \\
    10 & GSH068.5-03.0-043.3  &  11.1 &  0.63 &  12.9 &   0.81E-01 &   28.9 \\
    11 & GSH070.0-00.5-027.8  &  10.0 &  0.20 &   7.2 &   0.84E-04 &   15.9 \\
    12 & GSH077.0+02.5-039.2  &  10.7 &  0.17 &   6.2 &   0.13E-03 &   16.0 \\
    13 & GSH077.5+07.0-016.5  &   9.4 &  0.12 &   4.1 &   0.79E-06 &   17.3 \\
    14 & GSH083.0+03.5-053.6  &  11.6 &  0.19 &   4.1 &   0.28E-03 &   27.3 \\
    15 & GSH083.5+01.0-023.7  &   9.7 &  0.27 &   8.8 &   0.11E-03 &   17.9 \\
    16 & GSH085.5-02.0-024.7  &   9.8 &  0.17 &   5.2 &   0.11E-04 &   18.9 \\
    17 & GSH088.0-08.5-019.6  &   9.5 &  0.12 &   5.2 &   0.51E-05 &   13.9 \\
    18 & GSH088.0+01.0-084.5  &  14.3 &  0.31 &   4.6 &   0.66E-02 &   38.8 \\
    19 & GSH088.5+01.0-097.9  &  16.0 &  0.51 &   6.2 &   0.59E-02 &   48.4 \\
    20 & GSH089.0+01.5-115.4  &  18.7 &  0.62 &   8.8 &   0.29E-02 &   41.7 \\
    21 & GSH091.0-05.0-059.8  &  12.0 &  0.48 &  29.9 &   0.13E+01 &    9.4 \\
    22 & GSH092.0+02.5-023.7  &   9.7 &  0.27 &   6.7 &   0.98E-04 &   23.4 \\
    23 & GSH092.5-14.5-024.7  &   9.8 &  0.44 &  18.0 &   0.61E-02 &   14.2 \\
    24 & GSH093.5+18.0-017.5  &   9.5 &  0.32 &  18.0 &   0.26E-02 &   10.5 \\
    25 & GSH094.0+10.5-011.3  &   9.1 &  0.09 &   7.2 &   0.14E-04 &    7.4 \\
    26 & GSH096.5-03.5-028.9  &  10.0 &  0.13 &   4.1 &   0.11E-04 &   18.3 \\
    27 & GSH096.5+03.5-113.4  &  18.4 &  1.23 &  10.8 &   0.27E-01 &   66.6 \\
    28 & GSH097.0-11.5-011.3  &   9.1 &  0.09 &   6.7 &   0.15E-04 &    7.5 \\
    29 & GSH099.5-11.0-022.7  &   9.7 &  0.17 &   7.7 &   0.11E-03 &   12.6 \\
    30 & GSH100.5-02.0-038.1  &  10.6 &  0.16 &   6.2 &   0.14E-03 &   15.0 \\
    31 & GSH100.5+03.0-029.9  &  10.1 &  0.27 &  10.8 &   0.12E-02 &   14.6 \\
    32 & GSH102.0+08.0-017.5  &   9.4 &  0.34 &  17.5 &   0.66E-02 &   11.5 \\
    33 & GSH104.5-02.0-046.4  &  11.1 &  0.14 &   6.2 &   0.29E-03 &   13.2 \\
    34 & GSH105.0-08.0-013.4  &   9.3 &  0.28 &  23.2 &   0.14E-01 &    7.1 \\
    35 & GSH105.0-04.0-057.7  &  12.0 &  0.18 &   4.6 &   0.70E-03 &   23.1 \\
    36 & GSH106.5+13.0-043.3  &  11.0 &  0.11 &   5.7 &   0.12E-03 &   11.4 \\
    37 & GSH107.0+00.5-079.4  &  14.2 &  0.25 &   4.1 &   0.64E-02 &   35.8 \\
    38 & GSH107.5+01.5-034.0  &  10.4 &  0.24 &   7.2 &   0.62E-03 &   19.6 \\
    39 & GSH108.5+06.5-035.0  &  10.5 &  0.15 &   7.2 &   0.27E-03 &   12.0 \\
    40 & GSH109.0-08.0-060.8  &  12.4 &  0.20 &   6.7 &   0.41E-02 &   17.7 \\
    41 & GSH109.5-04.5-063.9  &  12.6 &  0.20 &   5.2 &   0.24E-02 &   22.4 \\
    42 & GSH109.5+17.5-018.6  &   9.6 &  0.13 &  10.3 &   0.35E-03 &    7.4 \\
    43 & GSH110.5+08.5-047.4  &  11.4 &  0.17 &   7.7 &   0.14E-02 &   13.0 \\
    44 & GSH111.5+01.0-079.4  &  14.4 &  0.29 &   5.2 &   0.12E-01 &   33.0 \\
    45 & GSH111.5+01.5-101.0  &  17.5 &  0.25 &   5.2 &   0.30E-03 &   27.9 \\
    46 & GSH117.0-04.0-026.8  &  10.1 &  0.15 &   6.7 &   0.27E-03 &   13.5 \\
    47 & GSH118.5+06.5-104.1  &  19.4 &  0.75 &   6.7 &   0.22E-02 &   65.7 \\
    48 & GSH119.5+04.0-095.8  &  17.8 &  0.50 &   7.7 &   0.45E-02 &   37.6 \\
    49 & GSH121.5-05.0-037.1  &  10.9 &  0.35 &  20.1 &   0.82E-01 &   10.2 \\
    50 & GSH121.5+14.5-057.7  &  12.9 &  0.17 &   6.2 &   0.38E-02 &   15.7 \\
    51 & GSH123.5-10.5-055.7  &  12.7 &  0.20 &   7.2 &   0.78E-02 &   16.6 \\
    52 & GSH123.5+06.5-021.6  &   9.8 &  0.59 &  29.9 &   0.42E+00 &   11.7 \\
    53 & GSH125.5-08.5-014.4  &   9.4 &  0.06 &   4.6 &   0.19E-04 &    8.3 \\
    54 & GSH127.5+00.5-105.1  &  22.6 &  1.01 &   8.8 &   0.28E-02 &   67.7 \\
    55 & GSH128.5-15.5-026.8  &  10.4 &  0.19 &  12.4 &   0.54E-02 &    9.1 \\
    56 & GSH128.5+03.5-033.0  &  10.7 &  0.08 &   4.1 &   0.46E-04 &   11.7 \\
    57 & GSH129.5+15.0-017.5  &   9.7 &  0.08 &   7.2 &   0.15E-03 &    6.8 \\
    58 & GSH129.5+06.5-067.0  &  14.6 &  0.31 &  10.8 &   0.13E+00 &   16.8 \\
    59 & GSH130.5+04.5-051.5  &  12.7 &  0.20 &   9.3 &   0.16E-01 &   12.5 \\
    60 & GSH131.5-13.0-021.6  &  10.0 &  0.07 &   5.7 &   0.61E-04 &    7.2 \\
    61 & GSH135.5-08.5-034.0  &  11.2 &  0.49 &  26.8 &   0.80E+00 &   10.7 \\
    62 & GSH136.0+07.0-030.9  &  11.0 &  0.15 &   9.3 &   0.26E-02 &    9.4 \\
    63 & GSH136.0+05.5-073.2  &  17.1 &  0.25 &   4.6 &   0.67E-03 &   31.1 \\
    64 & GSH139.0-03.0-069.0  &  17.0 &  0.84 &  11.3 &   0.13E+00 &   43.7 \\
    65 & GSH143.0+10.5-031.9  &  11.5 &  0.26 &  16.0 &   0.74E-01 &    9.5 \\
    66 & GSH144.0+02.5-050.5  &  14.5 &  0.54 &  13.9 &   0.12E+01 &   22.8 \\
    67 & GSH145.0-03.5-010.3  &   9.5 &  0.08 &   5.2 &   0.70E-04 &    8.9 \\
    68 & GSH145.0-10.0-063.9  &  18.4 &  0.30 &   4.6 &   0.53E-03 &   38.0 \\
    69 & GSH147.5-09.0-038.1  &  13.1 &  0.15 &   4.6 &   0.17E-02 &   18.8 \\
    70 & GSH150.0-12.0-017.5  &  10.4 &  0.05 &   4.1 &   0.27E-04 &    7.2 \\
    71 & GSH151.0-04.0-049.5  &  16.5 &  0.67 &  16.0 &   0.50E+00 &   24.7 \\
    72 & GSH155.0+13.0-016.5  &  10.6 &  0.17 &  10.8 &   0.76E-02 &    9.5 \\
    73 & GSH156.0-05.5-058.7  &  27.0 &  0.64 &   5.7 &   0.27E-03 &   66.8 \\
    74 & GSH159.0-00.5-053.6  &  27.9 &  0.82 &   4.1 &   0.15E-03 &  117.5 \\
    75 & GSH160.5+06.5-042.3  &  20.9 &  0.64 &   6.2 &   0.29E-02 &   60.6 \\
    76 & GSH164.0+00.5-016.5  &  12.1 &  0.20 &   6.2 &   0.53E-02 &   19.2 \\
    77 & GSH167.0-10.5-017.5  &  13.6 &  0.35 &   4.1 &   0.84E-02 &   50.1 \\
    78 & GSH168.0+11.5-023.7  &  19.4 &  0.36 &   8.2 &   0.62E-02 &   25.8 \\
    79 & GSH200.0-19.5+020.6  &  12.5 &  0.25 &   7.7 &   0.27E-01 &   19.0 \\
    80 & GSH200.5+06.0+022.7  &  12.5 &  0.14 &   5.2 &   0.23E-02 &   15.5 \\
    81 & GSH203.0-04.5+014.4  &  10.5 &  0.12 &   6.7 &   0.25E-02 &   10.7 \\
    82 & GSH203.0-38.5+013.4  &  11.0 &  0.09 &   4.1 &   0.36E-03 &   12.3 \\
    83 & GSH204.0-10.5+020.6  &  11.5 &  0.13 &   4.1 &   0.90E-03 &   17.8 \\
    84 & GSH206.0-19.5+021.6  &  11.4 &  0.11 &   4.1 &   0.76E-03 &   16.1 \\
    85 & GSH210.0-20.0+020.6  &  10.9 &  0.23 &  13.9 &   0.12E+00 &    9.7 \\
    86 & GSH210.5-04.5+029.9  &  12.0 &  0.18 &   6.2 &   0.74E-02 &   16.7 \\
    87 & GSH214.0+00.0+017.5  &  10.1 &  0.12 &   8.2 &   0.48E-02 &    8.6 \\
    88 & GSH215.5-03.0+053.6  &  15.2 &  0.17 &   6.2 &   0.45E-02 &   16.4 \\
    89 & GSH217.0+02.0+061.8  &  16.7 &  0.23 &   4.6 &   0.22E-02 &   29.6 \\
    90 & GSH217.0-06.0+026.8  &  11.0 &  0.17 &  14.9 &   0.99E-01 &    6.8 \\
    91 & GSH218.5-09.0+029.9  &  11.2 &  0.11 &   6.2 &   0.33E-02 &   10.8 \\
    92 & GSH219.5-15.0+015.5  &   9.8 &  0.12 &   8.8 &   0.56E-02 &    8.2 \\
    93 & GSH221.5+05.5+012.4  &   9.5 &  0.09 &   5.7 &   0.58E-03 &    8.9 \\
    94 & GSH222.0-05.0+025.8  &  10.6 &  0.11 &   9.8 &   0.10E-01 &    6.4 \\
    95 & GSH222.0-03.0+046.4  &  12.9 &  0.27 &   5.7 &   0.15E-01 &   28.4 \\
    96 & GSH223.0+03.0+069.0  &  16.5 &  0.28 &   5.2 &   0.39E-02 &   31.7 \\
    97 & GSH224.5-00.5+023.7  &  10.3 &  0.22 &  12.4 &   0.81E-01 &   10.2 \\
    98 & GSH226.5-09.5+031.9  &  10.9 &  0.30 &  22.2 &   0.12E+01 &    7.9 \\
    99 & GSH227.0-13.5+012.4  &   9.4 &  0.03 &   5.7 &   0.89E-04 &    3.4 \\
   100 & GSH228.5+02.5+066.0  &  14.6 &  0.22 &   5.7 &   0.55E-02 &   22.9 \\
   101 & GSH235.0+05.5+028.9  &  10.4 &  0.06 &   4.1 &   0.38E-03 &    8.8 \\
   102 & GSH238.5-02.0+068.0  &  13.8 &  0.18 &   4.1 &   0.17E-02 &   25.0 \\
   103 & GSH243.5-02.5+043.3  &  11.2 &  0.75 &  31.4 &   0.38E+02 &   14.0 \\
   104 & GSH243.5-15.5+025.8  &  10.0 &  0.16 &  11.3 &   0.68E-01 &    8.5 \\
   105 & GSH244.0-05.5+070.1  &  13.6 &  0.18 &   4.1 &   0.19E-02 &   25.7 \\
   106 & GSH244.0-09.0+035.0  &  10.6 &  0.21 &   7.7 &   0.44E-01 &   15.7 \\
   107 & GSH247.0-24.0+015.5  &   9.5 &  0.13 &   7.2 &   0.74E-02 &   10.4 \\
   108 & GSH250.0+01.5+067.0  &  12.9 &  0.20 &   4.6 &   0.46E-02 &   25.5 \\
   109 & GSH251.0-07.0+048.4  &  11.4 &  0.14 &   5.7 &   0.93E-02 &   14.4 \\
   110 & GSH252.0-02.5+034.0  &  10.4 &  0.08 &   5.2 &   0.25E-02 &    8.6 \\
   111 & GSH252.5+01.0+017.5  &   9.5 &  0.10 &  10.3 &   0.19E-01 &    5.6 \\
   112 & GSH253.0-08.0+013.4  &   9.3 &  0.06 &   4.6 &   0.56E-03 &    8.2 \\
   113 & GSH254.0+05.5+021.6  &   9.7 &  0.12 &   8.8 &   0.23E-01 &    7.8 \\
   114 & GSH254.5+00.0+056.7  &  11.9 &  0.31 &  10.8 &   0.25E+00 &   16.6 \\
   115 & GSH256.0-04.5+079.4  &  14.0 &  0.73 &  26.3 &   0.53E+01 &   16.4 \\
   116 & GSH256.5-07.5+035.0  &  10.4 &  0.19 &  13.4 &   0.36E+00 &    8.2 \\
   117 & GSH258.0+09.5+038.1  &  10.6 &  0.55 &  30.4 &   0.37E+02 &   10.6 \\
   118 & GSH259.0+03.0+016.5  &   9.3 &  0.08 &   6.7 &   0.56E-02 &    7.4 \\
   119 & GSH259.5-01.5+021.6  &   9.6 &  0.09 &   4.1 &   0.24E-02 &   13.2 \\
   120 & GSH260.5-04.5+064.9  &  12.5 &  0.18 &   4.1 &   0.34E-02 &   26.0 \\
   121 & GSH261.0+02.5+030.9  &  10.1 &  0.14 &   7.2 &   0.38E-01 &   11.0 \\
   122 & GSH262.5-01.5+041.2  &  10.7 &  0.38 &  19.6 &   0.56E+01 &   11.5 \\
   123 & GSH265.0-05.5+082.4  &  14.1 &  0.41 &  10.3 &   0.76E-01 &   23.2 \\
   124 & GSH266.0+04.0+037.1  &  10.5 &  0.12 &   5.2 &   0.95E-02 &   13.8 \\
   125 & GSH266.5-04.5+098.9  &  16.1 &  0.60 &   6.2 &   0.12E-01 &   56.8 \\
   126 & GSH266.5+07.5+020.6  &   9.6 &  0.18 &  12.9 &   0.52E+00 &    8.3 \\
   127 & GSH267.0-01.0+077.3  &  13.6 &  0.44 &   8.8 &   0.77E-01 &   29.5 \\
   128 & GSH267.5-03.0+114.4  &  18.5 &  0.73 &   8.2 &   0.17E-01 &   52.2 \\
   129 & GSH268.0-04.0+030.9  &  10.1 &  0.11 &   4.1 &   0.41E-02 &   16.1 \\
   130 & GSH269.5-07.0+077.3  &  13.6 &  0.24 &   7.2 &   0.13E-01 &   19.7 \\
   131 & GSH270.0-02.5+042.3  &  10.8 &  0.35 &  22.7 &   0.59E+01 &    9.0 \\
   132 & GSH270.0-05.5+096.9  &  15.8 &  0.35 &   4.1 &   0.12E-02 &   49.3 \\
   133 & GSH271.5-02.0+103.1  &  16.7 &  0.36 &   4.6 &   0.13E-02 &   45.8 \\
   134 & GSH272.0+04.0+024.7  &   9.8 &  0.23 &  18.5 &   0.13E+01 &    7.3 \\
   135 & GSH273.0+06.0+014.4  &   9.3 &  0.10 &   5.7 &   0.97E-02 &   10.4 \\
   136 & GSH273.5-07.5+031.9  &  10.1 &  0.41 &  29.4 &   0.15E+02 &    8.2 \\
   137 & GSH276.0+00.0+036.1  &  10.4 &  0.43 &  20.6 &   0.55E+01 &   12.4 \\
   138 & GSH278.5-02.5+026.8  &   9.9 &  0.19 &   7.7 &   0.46E-01 &   14.8 \\
   139 & GSH280.5-05.0+071.1  &  13.1 &  0.57 &   9.3 &   0.15E+00 &   36.0 \\
   140 & GSH281.5-01.5+085.5  &  14.6 &  0.61 &   6.7 &   0.23E-01 &   53.6 \\
   141 & GSH285.5-02.5+083.5  &  14.6 &  0.71 &   9.3 &   0.76E-01 &   45.3 \\
   142 & GSH292.0-01.5+050.5  &  11.6 &  0.38 &   9.8 &   0.13E+00 &   23.1 \\
   143 & GSH292.5-01.0+083.5  &  14.9 &  0.27 &   4.1 &   0.72E-03 &   38.4 \\
   144 & GSH297.0-00.5+069.0  &  13.6 &  0.95 &  15.5 &   0.65E+00 &   36.3 \\
   145 & GSH300.0-03.0+082.4  &  15.6 &  0.46 &   4.1 &   0.18E-02 &   66.2 \\
   146 & GSH304.0-01.0+061.8  &  13.3 &  0.58 &  12.4 &   0.95E-01 &   27.7 \\
   147 & GSH310.0-01.5+061.8  &  14.0 &  0.54 &   7.7 &   0.15E-01 &   41.1 \\
   148 & GSH319.0-03.5+023.7  &  10.4 &  0.43 &  11.3 &   0.80E-02 &   22.4 \\
   149 & GSH331.0-08.5+011.3  &   9.7 &  0.88 &  11.9 &   0.38E-02 &   43.8 \\
   150 & GSH333.5+11.5+014.4  &  10.2 &  0.90 &   7.7 &   0.19E-02 &   68.6 \\
   151 & GSH338.5-10.0+012.4  &  10.3 &  0.84 &   6.2 &   0.10E-02 &   80.2 \\

\end{longtable}
}

\end{appendix}

\end{document}